Title:  The variability of crater identification among expert and community crater analysts

Stuart J. Robbins*

Laboratory for Atmospheric and Space Physics; University of Colorado at Boulder, 3665 Discovery Dr., Boulder, CO 80309, United States

stuart.robbins@colorado.edu

*Corresponding Author

Irene Antonenko

Planetary Institute of Toronto, 197 Fairview Ave. Toronto, ON   M6P 3A6, Canada   and University of Western Ontario, 1151 Richmond St. N. London, ON  N6A 5B7, Canada

irene.antonenko@utoronto.ca

Michelle R. Kirchoff

Southwest Research Institute; 1050 Walnut Street, Suite 300, Boulder, CO 80302, United States

Clark R. Chapman

Southwest Research Institute; 1050 Walnut Street, Suite 300, Boulder, CO 80302, United States

Caleb I. Fassett

Department of Astronomy, Mount Holyoke College, 50 College St., South Hadley, MA 01075, United States

Robert R. Herrick

Geophysical Institute, University of Alaska Fairbanks, Fairbanks, AK   99775-7320, United States

Kelsi Singer

Department of Earth and Planetary Sciences and McDonnell Center for the Space Sciences, Washington University in St. Louis, 1 Brookings Dr., Saint Louis, MO 63130, United States

Michael Zanetti

Department of Earth and Planetary Sciences and McDonnell Center for the Space Sciences, Washington University in St. Louis, 1 Brookings Dr., Saint Louis, MO 63130, United States

Cory Lehan

The Center for STEM Research, Education, and Outreach at Southern Illinois University Edwardsville, Edwardsville, IL 62025 IL, United States

Di Huang

The Center for STEM Research, Education, and Outreach at Southern Illinois University Edwardsville, Edwardsville, IL 62025 IL, United States

Pamela L. Gay

The Center for STEM Research, Education, and Outreach at Southern Illinois University Edwardsville, Edwardsville, IL 62025 IL, United States

Running Head:  Variability in Crater Measurement

Abstract:

The identification of impact craters on planetary surfaces provides important information about their geological history. Most studies have relied on individual analysts who map and identify craters and interpret crater statistics. However, little work has been done to determine how the counts vary as a function of technique, terrain, or between researchers. Furthermore, several novel internet-based projects ask volunteers with little to no training to identify craters, and it was unclear how their results compare against the typical professional researcher. To better understand the variation among experts and to compare with volunteers, eight professional researchers have identified impact features in two separate regions of the moon. Small craters (diameters ranging from 10 m to 500 m) were measured on a lunar mare region and larger craters (100s m to a few km in diameter) were measured on both lunar highlands and maria. Volunteer data were collected for the small craters on the mare. Our comparison shows that the level of agreement among experts depends on crater diameter, number of craters per diameter bin, and terrain type, with differences of up to ~±45%. We also found artifacts near the minimum crater diameter that was studied. These results indicate that caution must be used in most cases when interpreting small variations in crater size-frequency distributions and for craters $\lesssim 10$ pixels across. Because of the natural variability found, projects that emphasize many people identifying craters on the same area and using a consensus result are likely to yield the most consistent and robust information.

# 1. Introduction

Impact craters are among the most common and numerous features on planetary surfaces in the solar system. They have been used for decades in various studies, from understanding the dynamics of the solar system to being a "poor man's drill" by excavating through numerous rock layers. This research relies on a key assumption: Impact craters can be reliably identified. Many applications, especially age estimates (McGill, 1977), also rely on measurements of crater diameter. It is generally assumed that both identification and measurement are trivial, but limited studies have shown this to not always be true; variations in crater identification and diameter measurement on the order of ~10% between individuals using the same measuring technique have been found (*e.g.*, Gault, 1970; Greeley and Gault, 1970; Kirchoff *et al.*, 2011; Hiesinger *et al.*, 2012).

Gault (1970) had approximately 20 people identify 1.3 million craters using Zeiss particle counters (this device allows the operator to match a pre-set circle size or "size class" of projected light onto a photograph, prick a hole through the photograph at the crater center, and the diameter is automatically registered on the instrument's display). He concluded, "'Calibration' and continuous cross-checks of each individual's work indicate that crater counts by different persons generally agree and/or can be reported within ±20% … ." Greeley and Gault (1970) used the same technique and data to further describe the dispersion among researchers. Measurements were made by five individuals on a single image and showed good agreement for small craters but dispersion in the number of craters among the largest diameters (their Fig. 3). Greeley and Gault (1970) found less than ±20% deviation from the mean for counts of >100 craters in a given size class, "a value that probably represents an irreducible minimum deviation imposed by the subjectivity of the counting." This variation rose to ±100% for counts with <4–5 craters in a given size class. The authors emphasized that a single individual may perform more consistent counts, but individual biases and differences from one day to the next – indeed, one hour to the next – explain why multiple individuals identifying craters on the same terrain are likely to yield the most reliable results.

Kirchoff *et al*. (2011) provide a more recent comparison with three researchers (two expert, one novice without crater counting experience) from the same lab who used the same technique to identify, measure, and, in this case, classify craters by preservation state. They used *Lunar Reconnaissance Orbiter* Camera Wide-Angle Camera (LROC WAC) images of Mare Orientale. The two experienced analysts had counts that differed by 20–40% in a given diameter range, while the novice counter identified numerous features that are probably not craters, differing from the other two by >100% over some diameter ranges. They also had significant variation among the preservation states attributed to each crater, despite a relatively coarse four-point scale. This work showed that despite common thinking that crater counting is fairly easy and straightforward, there is a learning curve and an individual's crater counts should be discarded during the learning process. It also showed that even well defined crater morphologies may be difficult to classify uniformly.

Hiesinger *et al*. (2012) also focused on lunar craters, in their case using LROC Narrow-Angle Camera (NAC) images at approximately 0.5 m/px. They were interested in reproducible results for better understanding the lunar cratering flux and performed a single test with two experienced researchers who used the same technique on the same image. The Heisinger *et al.* (2012) team found an overall variation of only ±2% between their analysts, a dispersion significantly less than previous work.

What this brief review indicates is that while there has been some discussion in the literature about agreement between different researchers' crater identifications, (a) there has been no thorough discussion on researcher variability, (b) no published study discusses the variability when using different techniques for crater identification and measurement, (c) variation in crater morphology has not been discussed (*e.g.*, sub-km craters appear substantially different at NAC pixel scales when compared with multi-km craters at WAC pixels scales), and (d) expert results have not been extensively compared with how well untrained or minimally trained crater counters do with the identification and measurement process. Given the proliferation of internet crowd-sourcing projects that ask laypeople to help in the data-gathering process, this last point determines if the public can assist in crater counting and produce results that are approximately as reliable as the experts.

For these reasons, the following work was undertaken. Eight researchers, with six to fifty years of experience identifying craters, identified and measured the diameter ($D$) of craters on a segment of a NAC image. This same region was also analyzed by volunteer "citizen scientists" through CosmoQuest's Moon Mappers ("MM") project which facilitates volunteer identification and measurement of craters and other features that are being studied in a variety of lunar research projects (Robbins *et al.*, 2012; http://cosmoquest.org). In addition, the experts identified and measured craters on a WAC image that covers both lunar mare and highlands. The experts worked independently, with each researcher using their own preferred technique (in total, seven different methods were employed).

These methods and the counting locations are discussed in Section 2 along with terminology, our display techniques, and statistical tests in section 2.4. Section 3 describes steps taken to ensure that our analysis is based solely on how different people identify craters, including how experts varied (3.1), how well the volunteers compared with experts (3.2-3.3), and how our data reduction process may affect results (3.4-3.5). Section 4 describes the overall crater populations found in each image and the variation among experts and between experts and volunteers. Section 5 moves from the population of craters from Section 4 to how well experts and volunteers agreed on the measurements of individual craters. Section 6 is an analysis of how crater detection depended on preservation state. Section 7 describes artifacts that we found near the minimum crater diameter. Section 8 is a short discussion of likely reasons for differences among experts and between them and volunteers. Section 9 summarizes the work and discusses implications and conclusions. Appendix A provides additional details on each researcher's technique, Appendix B summarizes each researcher's experience in the field, and Appendix C more thoroughly discusses our data reduction methods.

## 2. Methodology

### 2.1. Images Used

This work was motivated in part by the need to determine how untrained "citizen scientists" compare with experts. The experts in this study were asked to identify and measure craters in the same LROC NAC image that has been most studied by MM volunteers: M146959973L (Fig. 1). A portion of this image has been viewed by every MM volunteer because it is used as a calibration image to assess how well each individual performs. The image has a solar incidence angle of 77°, meaning that useful shadows are present to enhance local topography for better feature identification. It is also of general interest because it contains the *Apollo 15* landing site. MM uses a 4107×6651-pixel sub-image of M146959973L that is centered on the *Falcon* lander. The experts in this study were given a sub-image 33% of this size (Fig. 1), 4107×2218 pixels, to maximize participation among busy scientists. This sub-image contains on the order of 1000 craters $D \geq 18$ pixels (the limit of the MM interface).

The second image in this study was an LROC WAC that encompasses both mare and highland areas to allow comparison of expert crater identification on the two main lunar terrain types. The 1311×2802-pixel selected portion of WAC M119455712M covers the southern margin of Mare Crisium and the neighboring rim/highlands to the south (Fig. 2). It has a solar incidence angle of 59°, which is on the boundary of what is considered ideal for crater identification (Wilcox *et al*., 2005; Ostrach *et al*., 2011; Robbins *et al*., 2012).

Each image was downloaded from the Planetary Data Systems (PDS), processed in the USGS's *Integrated Software for Imagers and Spectrometers* (*ISIS*) via standard radiometric and geometric techniques, projected to a local Mercator projection, exported as PNG files, and distributed to each researcher. For several of the crater identification and measurement techniques, GIS-ready files were required. To ensure uniformity, Robbins imported the images into *ArcMap* and exported the GIS-ready files, and he distributed them to the researchers, too.

### 2.2. Techniques and Personnel

Each researcher employed one or more different interfaces and methods, with Robbins using two interfaces and Antonenko using three to replicate their NAC crater counts; Antonenko used two for her WAC counts. The software and methods used are briefly described here and they are detailed in Appendix A. Tables A1-A2 summarize the numbers of craters found, methods used, and image manipulation applied by each researcher.

The most common software was ESRI's *ArcGIS* suite, with Antonenko, Fassett, Herrick, Robbins, Singer, and Zanetti using it. However, four different methods were employed. Robbins used native tools to trace crater rims and then his own software to automatically batch-fit circles to each traced rim. The other researchers used software extensions. Fassett and Zanetti used "CraterTools" (Kneissl *et al*., 2011) which uses a three-point method to determine a circle; Herrick also used this interface for his WAC dataset. Antonenko and Herrick (the latter for NAC data only) used the USGS's "Crater Helper Tools" which also applies a three-point circle identification technique (Nava, 2011). Singer used native tools to draw a chord through the crater center, from rim to rim, and the spherical length is calculated using Jenness Enterprise's "Tools for Graphics and Shapes" (Jenness, 2011).

The second-most common software application for experts was JMARS, produced and maintained by Arizona State University (http://jmars.asu.edu). This was used by Antonenko (second interface) and Kirchoff. JMARS has a built-in crater measurement tool that has two options for determining the diameter: three-point circle fits (Kirchoff) and a re-sizeable circle tool (Antonenko).

The third interface, used only by Chapman, was the Smithsonian Astrophysical Observatory's DS9 (not an acronym) visualization software with several add-on tools written by various researchers. For crater handling, the POINTS tool, developed at Cornell, was used which requires researchers to identify an even number of points along the crater rim. The researcher then enters a numerical crater preservation state to complete the identification, and a circle (or ellipse) is fit.

The fourth software was the Moon Mappers online interface, used by Antonenko (as her third interface), Robbins (as his second interface), and all lay volunteers; this software was only used for the NAC image. To identify a crater, the user clicks on (or near) the crater center and drags outwards, dynamically displaying a circle of that radius and center position. The circle is color-coded based on size: red if the diameter is $D < 18$ px and green when $D \geq 18$ px (red craters are considered too small for confident identification by MM volunteers). When the user is satisfied that the drawn circle matches the crater, they release the mouse. Green craters are saved and displayed (red craters are discarded), but they can still be moved, re-sized, and erased if desired. In MM, images are shown in 450×450 pixel subsections, but different images are at different "zoom" levels to allow crater identification over a large range of diameters. The 18-px cutoff was chosen to (a) limit the number of craters volunteers need to identify on any single image (avoid user fatigue), (b) the approximate limit to which lay persons can easily identify a crater (determined during development work at Southern Illinois University Edwardsville), (c) is two pixels smaller than 20-px, which was our original buffer, due to technical limitations at the time, and (d) ensures craters $D \gtrsim 10$ m will be identified when using average-resolution NAC images.

All expert researchers involved in this project have at least six years experience identifying and measuring craters, and one (Chapman) has over fifty years experience in this work. In contrast, the MM volunteers have no stated experience, and though Robbins and Antonenko both have marked a few craters for the MM project and so contributed to the "volunteer" dataset, it is safe to estimate that ≪1% of MM craters were marked by experts.

## 2.3. Cluster Analysis to Compare Craters and Create a Reduced Catalog

Before any analysis could be conducted, the various measurements had to be properly aligned and in the same units; we chose to perform all analyses in pixel space in order to be scale-independent (*i.e.*, all units are pixels and we compare NAC and WAC diameter-dependent results in pixels as opposed to meters and kilometers). Robbins', Chapman's, and all MM-marked craters were already in pixel space upon export from software. For all other data, crater diameters and locations were scaled linearly by the pixel scale (m·px$^{-1}$) output from *ISIS*. The NAC image was small enough and close enough to the equator that a non-linear transform was unnecessary. Non-pixel-space crater locations and diameters for the WAC image were corrected by cos(latitude) to convert to pixel space after the initial linear scaling.

After these corrections were applied and all craters were in pixel space (Figs. 1 (NAC) and 2 (WAC)), a clustering code was used to separately group the expert and volunteer markings into "reduced" crater catalogs for subsequent analysis. Most analyses presented in the next section require such a reduced catalog to determine how the markings varied for each crater.

We detail the development of the clustering code in Appendix C. In brief, the automated code was a modified two-dimensional DBSCAN (Density-Based Spatial Clustering of Applications with Noise) developed by Ester *et al.* (1996). The original code requires only two *a priori* inputs: a "reachability" parameter and the minimum number of points in a group required to be considered a valid cluster. Reachability is how cluster members are found, where a point is considered "reachable" by another point if it is within a certain distance of it or other members of the cluster to which it may belong.

The code was modified to incorporate a diameter-dependent scaling of distance and to include a second reachability parameter dependent solely on diameter. The former means that, for example, two 20-px-diameter craters 10 pixels away from each other would not be considered reachable and hence members of different clusters, but two 200-px-diameter craters 10 pixels away would be considered the same feature. The latter means that if a crater was reachable by another based on location, it also needed to be reachable based on diameter. This allows us to separate overlapping craters, *i.e.*, a small crater superposed on a larger crater that under a normal two-dimensional DBSCAN code would be grouped together into one feature.

The input data were crater location $\{x, y\}$ and diameter $D$ in addition to the confidence $c$. Confidence is a score from 0 to 1 used to rate how well each MM volunteer performs versus an expert (Robbins) on calibration images, and it is used to calculate weighted means of the clustered craters. For experts, we set $c = 1$. The data output for each cluster of markings are: (a) weighted mean $\bar{x}$ with sample standard deviation, $\delta x$; (b) weighted mean $\bar{y}$ with standard deviation, $\delta y$; (c) weighted mean $\bar{D}$ with standard deviation, $\delta D$; (d) number of points $N$ in that cluster; and (e) weighted mean of the confidence for the craters that went into that cluster, $\bar{c}$.

## 2.4. Terminology, Display Techniques, and Statistical Tests

In this section, we detail the comparison of expert crater markings and how those relate to volunteer results. After section 3, unless otherwise stated, all "catalog," "reduced," or "ensemble" expert data are from the clustered results of all expert data (11 total interface/expert combinations for NAC, 9 for WAC) where $N \geq 5$ persons (NAC) or $\geq 4$ persons (WAC) needed to find the crater for it to count; the data were clustered even if the diameters were smaller than a stated *a priori* cut-off (18 px for NAC), and the cut-off was performed on the post-clustered results. The volunteer data are the output from the clustering code with $N \geq 7$ persons having identified the crater, unless otherwise stated (images were viewed by 12–26 people). All error bars on crater populations are based on Poisson counting statistics ($N^{1/2}$ where $N$ is the number of craters found) unless otherwise stated. All error bars on other analyses are based on Gaussian sample standard deviations (square-root of the variance normalized by $N$) unless otherwise stated.

We present and analyze the crater measurements in both cumulative and relative size-frequency distributions (CSFD and R-plot; Crater Analysis Techniques Working Group, 1979). The CSFDs are shown as true cumulative counts without binning. We use the Kolmogorov-Smirnov (K-S) test, which determines if two measured distributions come from distinct distributions, to compare the identified craters' CSFDs. This test finds the maximum vertical separation between the two curves for the full range of *x*-values (we use crater diameter), normalized to a cumulative value of 1.0. This maximum is then compared with expected values to determine the probability (*P*-value) that the null hypothesis (the distributions are the same) is rejected. We used a *P*-value of >0.05 to accept the null hypothesis. A *P*-value of ≤0.05 is used to state that two populations are likely different, and a *P*-value of ≤0.01 is interpreted to reject the null hypothesis and state that they are different; 0.05 is the most common value used in statistics but the smaller value was used because the K-S test does not take error bars into consideration.

## 3. Results: Overall Crater Population: Separating Independent Variables Before Analysis

In this sub-section, we present a step-by-step analysis of the numbers of craters found and how well they match with individual researchers' results. We do this across all combinations of individuals and interfaces to determine where – if at all – artifacts or biases may arise that affect the later analyses.

### 3.1. How do the experts vary in their interfaces of choice?

One of the fundamental questions in this study is: What is the variation among experts and what role does interface play in this variation? These are motivated by a desire for the "right answer" about the number of craters on a given surface, and this has implications regarding uncertainty for studies such as the modeled surface ages. The top portion of Table 1 provides a summary of the numbers of craters found by each expert in the NAC data, listed by expert and interface, and separated into crater diameters $D \geq 18$, 20, 25, 30, 50, and 100 pixels. The next two portions of Table 1 show the mean, median, standard deviation, and relative standard deviation "$\sigma_R$" (standard deviation divided by the mean) among the experts for these different diameters for all interfaces except MM and then with the expert data from MM included (data from MM is treated separately at first to show whether it can accurately replicate results from the other interfaces). Table 2 shows the same information for the WAC image, separated into highlands and mare units, with no MM data (since that interface was not used in the WAC study). These data are also shown in Figs. 3 (NAC) and 4 (WAC) as a function of diameter.

NAC: The data indicate there is at least ±20% dispersion among experts in the number of craters found at any given diameter in the NAC data. In their interfaces of choice, the relative standard deviation ($\sigma_R$) is a minimum of 20.7% for $D \geq 18$ px craters and grows to 31.5% for $D$ $\geq 100$ px craters (Fig. 3). A hypothesis we test in Section 6 is whether this may be due to crater preservation state because large, degraded craters are numerous in this region of the Moon and at this pixel scale. Alternatively, we might conclude that experts converge better when a region is dominated by numerous small craters rather than a few large craters. This is similar to what Greeley and Gault (1970) found, with $\sigma_R$ converging to $\approx$20% once the number of craters found is $\gtrsim$20. When including the results of Antonenko and Robbins from the MM interface, the medians remain nearly identical while $\sigma_R$ decreases by 0.1% for the smallest craters and 3.3% for the largest craters. The cause is the MM expert crater identifications were near the median so this decreased the overall spread in the results.

Table 1 also shows three different techniques for determining the answer to "how many craters are there?" in this image: One could use (1) the mean or (2) median of the individual experts' results, or (3) the clustering algorithm's results. Comparing the first and second statistical measures shows that the median is slightly lower than the mean for smaller craters. This indicates that our sample of 11 expert datasets is not uniformly distributed but weighted towards fewer craters. For the third technique, the complication arises that the experts' results are not identical. The clustering code requires a threshold number of people who must identify a crater ($N_{threshold}$) for it to be considered "real" and counted. Changing $N_{threshold}$ alters the number of craters identified. To determine an $N_{threshold}$ that we considered to be a reliable consensus in the crater population amongst the expert analysts, we calculated the number of craters that were found in each diameter range using different values for $N_{threshold}$ (Table 1) and compared these values with the mean results from technique 1 above. The closest matches are boxed in Table 1. While there is no perfect $N_{threshold}$, the results indicate that regardless of whether or not the data from the experts' Moon Mappers interface are included, the $N_{threshold}$ value for a reasonable consensus is 5; *i.e.*, the crater must have been identified at least 5 times for it to be a member of our final dataset; this is indicated as bold in Table 1. Note that Fig. 3 shows results when normalizing the data both to the combined and normalized experts' results (technique 1) and clustered data (technique 3); it shows nearly the same values, indicating our choice of $N_{threshold} =$ 5 is reasonable for this application (see the Discussion for how varying this affects results and how different values may be appropriate for different applications).

WAC:  The data indicate there is at least ±20% dispersion among experts in the number of craters found at any given diameter in the WAC data except for $D \approx 10$ px craters in the mare region (Table 2, Fig. 4).  The overall trend from Fig. 4 is similar to the NAC data where experts converge on a similar crater count when the region is dominated by smaller craters; this holds until craters are smaller than the completeness level for all experts (the diameter to which we estimate that all craters were identified).  Determining $N_{threshold}$ for each WAC terrain followed the same technique as for NAC:  We calculated the number of craters that were found in each diameter range using different values for $N_{threshold}$ (Table 2) and compared these with the mean results from the experts.  While there is again no perfect $N_{threshold}$, the results (boxed values being the closest to the mean) indicate that a reasonable $N_{threshold}$ is 4; *i.e.*, the crater must have been identified at least 4 times for it to be a member of our final dataset.  In both the WAC and NAC cases, $N_{threshold} = \text{floor}(0.5 \cdot N_{experts})$.  While it is not possible to state this would hold in all cases, it suggests a starting point if similar work to this is done in the future, and future work may find that a fixed value is more appropriate (*e.g.*, $N_{threshold} = 4$).

## 3.2. Does the MoonMappers interface work as well as an expert's preferred interface?

To facilitate inter-method comparison, both Antonenko and Robbins identified craters from the NAC image in the MM interface.  The results are listed in Table 1 and illustrated in Fig. 5 using CSFDs and R-plots.  Robbins' results are nearly identical over the entire 18–450 px diameter range; they agree to within $0.5\sigma$ for $D > 30$ px, the results only deviate to $>1.0\sigma$ for craters $D < 21.5$ px, and some of this deviation is due to the rounding that MM performs on most crater diameters.  A K-S test on craters $D > 20$ px between the two datasets has a *P*-value of 0.054, indicating that the null hypothesis is accepted (they are the same distribution).  Antonenko's results are more complicated:  Craters from all three interfaces agree within $1\sigma$ for $D > 80$ px.  For $30 < D < 80$ px, data from ArcGIS diverges to fewer counts than the data collected in MM and JMARS.  In the range 25–30 px, the population of craters from the MM interface decreases relative to JMARS to match that from ArcGIS for $D < 25$ px to within $0.2\sigma$; this is after the effects of the rounding to the nearest pixel diameter in JMARS are considered.

On reflection, Antonenko thinks the lower population of ArcGIS-found craters in the ~30–80 px range is due to a conscious decision to not identify heavily degraded craters when counting in that interface; her decision changed when performing crater counts in JMARS and MM.  We explore this further in Section 6.  Setting this aside, the results from this analysis show that the Moon Mappers interface allows crater measurement at least as good as those achieved by experts using these three different techniques for crater identification over a broad range of crater diameters.  The exception is at the small end, where both Antonenko and Robbins found fewer craters when using MM.  This is explored further in Sections 3.3 and 7.

## 3.3. How well does an expert compare with volunteers in the Moon Mappers interface?

Previous sections demonstrate that experts' crater identifications generally agree to within ~20–30% when they utilize their interfaces of choice, and comparable agreement is found between experts' crater identifications from the MM interface and their preferred interface(s).  We address how well experts agree with each other in detail in Section 5.  The next question is how experts compare with volunteers when using the MM interface.  To answer this, the volunteer data first must be reduced (clustered) and a suitable $N_{\mathrm{threshold}}$ must be determined.  (Note: As of April 5, 2013, approximately 12–26 persons (mode = 14) have viewed each sub-image of the study area.)

The final part of Table 1 shows the same "method 3" discussed in Section 3.1 applied to volunteer data.  The results were more scattered than the experts', and craters $D < 25$ px were not used to set this threshold due to aliasing effects (see Section 7).  The volunteer data contains the fewest craters relative to experts in the ~25–40 px range which we hypothesize (and test in Section 6.1) is due to a larger proportion of heavily modified craters in that range. We find again that while no $N_{threshold}$ is ideal, setting $N_{threshold}$ = 7 gives the best comparison with expert data; this is approximately 60% of the number of persons who viewed each sub-image.

These resulting volunteer data are compared with Antonenko's and Robbins' MM interface-based data in Fig. 6. These data indicate that for a large diameter range ($30 \lesssim D \lesssim 500$ px), the ensemble volunteer data match the expert data at least as well as individual experts match each other. For craters nearing the limits of the MM interface ($18 \leq D \lesssim 25$ px), the number of craters found by volunteers relative to experts greatly increases until about 1–2 px larger than the 18 px cut-off, at which point there is a dramatic flattening of the CSFD (relatively few craters found). A K-S test of the three datasets show that for craters $D \geq 23$ px, all three datasets have the same population with a minimum $P$-value of 0.050 (Antonenko-Volunteers) and maximum 0.28 (Antonenko-Robbins). We conclude that the volunteers compare well with the experts when using the same interface except for the apparent artifact at small diameters. We explore this artifact in Section 7.

### 3.4. Does the clustering code affect results?

In the previous sub-sections, we have shown that not only do experts *generally* agree with each other across all interfaces, but they also agree with volunteer data (we perform more explicit tests in section 5). Now we ask whether the clustering code is appropriate for our analysis. So far, we have compared individuals' results with those of the volunteers' clustered ensemble. To determine whether the clustering algorithm can affect results, we took Antonenko's and Robbins' data from MM and copied them 15 times each and then multiplied the positions and diameters by random numbers to simulate variability. The random numbers were drawn from a $\mu = 0$, $\sigma = 0.1 \cdot D$ Gaussian distribution, such that smaller craters were given a smaller amount of scatter relative to larger ones. Antonenko's and Robbins' data were clustered separately and then compared with the original experts' ensemble result. As an additional test, Antonenko's and Robbins' data were duplicated seven times each, multiplied by the Gaussian noise, and clustered together. The various clustered datasets were then plotted together with the results of their original MM data, similar to Fig. 5 but not shown here. They were also visually inspected as in Fig. 1. The results show no appreciable difference: the clustered data agree to within $0.1\sigma$ of each expert's original MM results over most diameters in all three tests (Antonenko ×15, Robbins ×15, (Antonenko + Robbins) ×7). The only deviation from $0.1\sigma$ agreement was for $D < 30$ px in Robbins' data and $D > 150$ px in Antonenko's data (in both cases, for those diameters, this clustering test resulted in fewer craters; agreement was, however, well within $0.5\sigma$). When K-S tests were conducted to determine how similar the populations were, the $P$-value was $>0.05$ (they are the same population) for $D \geq 18$ px for Antonenko but only $D \geq 19$ px for Robbins. Together, these tests indicate that the clustering code does not introduce significant artifacts and is appropriate and useful for this type of data.

## 3.5. Does the clustering code act differently on expert versus volunteer data?

The final independent variable in data reduction before population results can be discussed is whether the clustering code behaves differently on expert data versus volunteer data. To answer this question, we took the results from the test in Section 3.4 and compared them with the volunteers' data (similar to Figs. 1 and 6). We found no difference. We then compared this to the ensemble (clustered) expert data and also found no appreciable difference.

We have shown through the analyses and tests throughout this sub-section that none of the independent variables (interfaces, personnel, and data reduction code) introduce significant biases or systematic errors (with the two caveats of crater preservation state and the smallest crater diameters). Ergo, we are confident these data are similar enough to compare directly in the subsequent sections, and any significant differences (save those two caveats) are uniquely a function of experts versus volunteers and are neither related to our data reduction nor the interfaces in which data were gathered.

## 4. Results: Overall Crater Population in WAC and NAC Images

When identifying craters on planetary surfaces, researchers typically assume they have identified all the craters on a surface, at least to within the standard Poisson counting uncertainty, but that from one moment to the next their results will not vary and that someone performing the same task will have similar results. Prior work has found this to *not* be the case (Gault, 1970; Greeley and Gault, 1970), and this work re-emphasizes that point: Even among crater experts, "the number" of craters mapped on a surface can vary by a significant percentage.

We have examined this in multiple ways on two types of surfaces (mare and highlands) with morphologically and morphometrically different craters (NAC and WAC). First, we examined the numbers of craters identified at certain diameters (NAC in Table 1, WAC in Table 2): we present those results visually in $2^{1/8}D$ multiplicative intervals in Figs. 3 and 4. The data from NAC craters show that over a broad range of diameters, individual researchers may vary in the number of craters identified by up to ±40% from the mean (when the number of craters identified is <10) with a diameter-dependent $\sigma_R$ that is ≈20% for $D \approx 20$ px and grows to ≈30% for $D = 100$–200 px.

The nature of the terrain in the NAC data likely plays a role in the variability between experts. Craters of all sizes show a range of modification states (see Section 6.1). This, combined with the measured density of craters, suggest the count region is in saturation equilibrium for most of the crater sizes we measured, consistent with expectations for the lunar maria (*e.g.*, Shoemaker, 1965). As such, this surface may be a "worst case" scenario for the repeatability of crater counts, although we show later that the lunar highlands fare more poorly in terms of expert reproducibility (though it is possible a higher sun angle in the WAC image also contributed to poorer repeatability).

The counts in the WAC study area exhibit different behavior on the two different terrain types that were measured. Mare crater counts have a dispersion of ±50% when $N_{craters} \approx 5$ that rapidly shrinks to a minimum of ±13% at $D \approx 10$ px, when $N \approx 80$. This is more of a "best case" scenario for crater statistics since ~10-px craters (~700 m) in WAC data are too large to have significantly degraded over the age of the maria, but they are numerous enough to provide useful statistics that minimize the variations between workers.  At smaller sizes, the dispersion between researchers increases to ±30–40% ($N \approx 800$), but this is below the size where researchers were complete in their counts ($D \approx 4$–7 px).

Highlands counts on the WAC data also showed a large dispersion for $N_{craters} < 10$ ($D > 100$ px), but the dispersion then settled between approximately 35–40% over the larger $5 < D < 100$ px range.  The generally size-independent dispersion versus numbers of craters found in the highlands is indicative of poor agreement among the researchers due to the highly modified surface and high sun angle.

From these data, attempting to converge on a single best method for combining the measurements to derive the "best" result for the number of craters identified is difficult.  In Section 3.1, we discussed three different methods:  Mean or median of the expert results at any given diameter or clustering.  While these results show that there is no single correct answer, we can at least identify an optimum method and result.  Another way of examining this is to look at the populations of the craters in a CSFD and R-plot.  Fig. 7 shows these for all surfaces and images examined.  In addition to the visual inspection, we quantified the agreement between each expert, the clustered experts, and MM volunteers (NAC-only) by using the K-S test.  For this discussion, we focus on diameters ≥25 pixels for NAC and ≥7 pixels for WAC due to issues with smaller diameters (see Section 7).

NAC:  K-S tests were run for the 91 permutations comparing each expert, the expert ensemble results (with and without expert MM data), and the volunteer ensemble results for $D \geq 25$ px.  We find the majority of datasets match well with the majority of others (55 comparisons had $P$-values >0.05, 23 had 0.01–0.05, and only 13 of the 91 had $P<0.01$).  Those experts who agreed least well with others were Singer, Kirchoff, and Antonenko (when using the JMARS interface).  Kirchoff's and Antonenko's JMARS results agree very well with each other ($P$-value = 0.6) but not with others.  We think this is likely an artifact caused by a limitation in a version of JMARS that limited crater diameters (in meters) to integer values (this limitation was removed in the early April 2013 release).  When a small amount of random scatter is added to their diameters (on the order of 0.5 meters) to remove this rounding effect and the K-S test run multiple times as a mini Monte Carlo experiment, their data agree much better with the rest of the experts. This leaves Singer's results as the primary outlier. Anticipating differences, each expert was asked to thoroughly describe their method of identifying craters. Singer was unique (except for Antonenko's work in *ArcMap*) in that she excluded craters that were highly modified. As we discuss in Section 6, the majority of craters $D \gtrsim 30$ px are heavily modified on the saturated mare surface in this NAC. This presumably reduced Singer's total crater count, which was the smallest for $D \geq 25$, and affected the larger craters more, resulting in an apparent different population.

WAC: The K-S test was run for the 45 permutations of each expert and the ensemble results for $D \geq 5$, 6, 7, 8, 9, and 10 px each for both the mare and highlands regions. As would be expected visually from both Figs. 4 and 7, the mare results show highly consistent populations for all persons for $D \geq 9$ px, and for everyone except Chapman for $D \geq 8$ px and Herrick for $D \geq 7$ px (this is consistent with Herrick's self-described completeness estimate of only $D \geq 9$ px and Chapman's estimate for $D \geq 12$ px (see section 7.4)). One stray very large crater found by Robbins is likely an artifact of the limited context for the image and is likely just a scallop in the rim of Mare Crisium; this shows that broader context can be useful in crater identifications. The highlands data are considerably more complicated with much less agreement. The CSFDs in Fig. 7 and Table 2 show that Zanetti found the fewest large ($D > 50$ px) craters while Singer found the fewest overall, although again Singer did not focus on identifying highly modified craters. Chapman found the most $D \geq 10$ px craters by a factor of 64% over the next-most (Robbins). Population-wise, the overall R-plot and CSFD slopes look similar except for (a) Kirchoff being steeper for $D < 8$ px, (b) Herrick is much shallower for $D < 15$ px (discussed more in section 7.4), and (c) Chapman's craters increase at a slightly greater rate than others' through his estimated completeness around $D \approx 10$–12 px. This is reflected in the K-S test results where all persons only agree with each other and the ensemble for $D \geq 14$ px, and Herrick disagrees the most with the others for smaller diameters. *Post hoc*, Herrick thinks his deficit was because he is less familiar with lunar terrains and more conservative than the rest of the experts in this study in identifying somewhat circular irregular terrain as a degraded impact crater. From this, we can easily say that the lunar highlands' crater counts are most prone to subjectivity, and we discuss the implications of this in Section 9.

In this analysis of the crater population, we also wanted to answer the question: Is there a variation between experts on different terrain types? We can use the above-discussed data to answer that question because the WAC data are separated into mare and highlands while the NAC mare craters are morphologically distinct from those at WAC pixel scales. The WAC mare, NAC, and WAC highlands is the order for most to least agreement, and this also generally reflects the level of qualitative ease the experts assessed for each image and region (Table A1). They generally considered the mare "clean" with fewer ambiguous craters at the WAC scale. The NAC had mixed assessment where some considered it straightforward while others considered it more difficult; the most difficulty was the large number of highly modified craters ~10s px in diameter. The general consensus was that the highlands were difficult, several experts indicated that elevation data would have helped (and perhaps be a separate, future study), and several others requested guidance on the larger or more ambiguous somewhat circular depressions (in the interest of remaining as uniform and blinded as possible, Robbins gave everyone the same instructions, even when prompted for more: "If you would normally identify this feature as a crater, then do so; if not, then don't."). In the end, the conclusion from this limited study of three "terrain" types with one example of each, we can answer in the affirmative that there is a variation among experts on different terrain types. Perhaps as consolation, we can also say that at least those who participated were generally aware of the ambiguity and relative uncertainty in their overall crater counts on each type.

## 5. Results: Examining the Agreement on Individual Craters

The purpose of this sub-section is to answer the question: How do identifications of individual craters compare between experts and volunteers? For this question, the overall number of craters identified is not a factor. To answer it, we performed two separate analyses. The first was to examine the standard deviations ($\sigma$ from the mean) of the locations and diameters of the different markings included in each crater cluster. The second was to compare individual clustered craters from the experts' and volunteers' datasets and determine if there is a systematic offset between the two's clustered diameters and/or locations.

For both tests, a brief preamble on the number of identifications ($N$) that went into each crater is necessary. For the NAC image, 1375 "craters" with $1 \leq N \leq 12$ were identified from the 11 expert datasets. There were two $N = 12$ craters, and both of these were considered acceptable to leave in the reduced dataset: In one case, an expert had identified two very closely overlapping similarly sized craters but all others identified them as one crater, and the other case was where one expert identified one very similarly-sized crater inside another, but all other experts identified them as one crater. Of the 889 craters with $N \geq 5$ markings, 450 (51%) were identified by *all* experts in *all* interfaces. An additional 15% were found by 10 of the 11, and 10% by 9 of 11, such that at least 9 expert markings were included in 75% of all the craters in the final "expert catalog." Ergo, despite setting the NAC threshold at $N \geq 5$, only 25% of the craters had 5–8 markings. While the WAC data from the experts have different absolute numbers for this analysis, the percentages are similar. For the volunteer catalog, the $N$ distribution was much different. There were 3045 "craters" identified with $1 \leq N \leq 62$ (this large number is possible because sub-images overlap and are at different zoom levels). Thirty-three percent of the final catalog of 813 craters (from $N \geq 7$) were found by $N \geq 14$ persons, 50% were found by $N \geq 12$ persons, and the percentage rises by 10% for each additional individual removed from the threshold. This implies a significant difference exists between experts and volunteers: While all experts found the majority of craters, there was a monotonic decrease with an exponential tail in the number of volunteers who found each crater in the final volunteer catalog.

The first analysis in this sub-section, comparing the standard deviation of identified craters' locations and diameters ($\delta x$, $\delta y$, $\delta D$), is shown in Fig. 8 as a function of average crater diameter. For the NAC data, the experts' spread in location shows no significant dependence on diameter, and the standard deviation in the ensemble positions averages about 5±2%·$D$. The deviation in diameters shows a weak dependence expressed as $\delta D = 0.14 - 0.22 \cdot D^{-0.39}$, starting around 7±2% at $D \approx 20$ px and climbing to about 10±4% for $D \approx 200$ px. The WAC results are somewhat different, where in all cases (location and diameter for mare and highlands) the dispersion is approximately 10±3% for $D \approx 7$ px and falls for larger craters. The diameter dispersion of highlands craters drop the least to a non-statistically different value of about 7±5% by $D \approx 40$ px. The locations of the highlands craters fall the most to 4±2% by the same diameter. The locations and diameters of mare craters follow the highlands craters' locations, but there are not enough craters per bin for $D > 20$ px to derive any meaningful results. There is no significant difference in $\delta x$ from $\delta y$ in the NAC and WAC data.

In contrast with the expert results for NAC, the volunteers show a comparable amount of scatter in their identifications of crater location and diameter for $D \approx 20$ px, but this quickly climbs to an approximate equilibrium of $\approx$±20% in crater diameter and ±10% in crater location for $D \gtrsim 50$ px. The comparable scatter at small diameters may be because the MM interface has a fixed pixel scale and volunteers are forced to only include craters $D \geq 18$ px, and so they may be more careful around that minimum diameter. It may also be because craters smaller than 18 px that may contribute to that minimum size bin and a scatter within it cannot be identified. In addition, there was a small but consistent offset where $\delta x$ was ~1–2% larger than $\delta y$ when normalized to crater diameter. This might be an indicator of a psychological effect where it is easier to precisely locate mouse positions in the $y$ direction rather than the $x$ direction and could be an area of future computer interface research. It could also be an indication of how lighting angle affects shadows which could influence untrained volunteers more than experts.

To test the idea that there is a minimum dispersion in the smallest MM craters because those $D < 18$ px are removed, we pre-clipped all NAC expert data with diameters $D < 18$ px, re-clustered them, and performed the same analysis as in Fig. 8 (not shown). We find that the effect seen in the volunteer data is duplicated, but to a smaller extent, and it only affects crater diameters and not location (the smallest bin with crater location standard deviations changes from 6.1±2.2% to 6.0±2.0%). Unexpectedly, the smallest three diameter bins are affected instead of just the smallest, though the smallest is the only statistically significant change, from 7.0±1.8% to 3.8±1.0% (larger bins were affected because the diameter "reachability" parameter could still include a $D \approx 17$ px crater even in a $D \approx 22$ pixel cluster). The conclusion is that this effect can account for a lot of the difference from larger craters at the smallest diameters, but accounting for the offset at diameters up to $D = 50$ px requires additional explanation and is not an artifact of the 18 px cut-off. From this comparison overall, we conclude that, as a whole, experts are more consistent from person-to-person in crater measurement and identification, where the scatter in diameter is slightly greater than position but averages around the 5–10% level, while volunteers average around ±10% for location and ±20% for diameter; also, only weak diameter-dependent effects were found for both volunteers and experts in NAC data.

The second comparison method required matching the reduced expert craters with the reduced volunteer craters and so was only done for NAC data. Clearly, there was not a one-to-one comparison here because the number of craters in each was 889 and 813, respectively, but there were 750 unambiguous matches between the two. Note that there are only 699 matches when both datasets are limited to $D \geq 18$ px (*i.e.*, the additional 51 are for expert craters $D < 18$ px matching volunteer craters $D \geq 18$ px); the diameters of these craters are displayed in Fig. 9. A correlation coefficient calculation indicates a very high 1:1 correlation (0.995) between the two.

# 6. Results: Dependence on Crater Preservation State

Craters typically form with sharp rims, deep cavities, an ejecta blanket, and other morphologies that make them relatively easy to identify. As craters age, they are eroded, infilled, and resurfaced; these processes act to mute the sharper features and make identification and measurement more difficult, for it is the sharp transition from light to dark in a circular pattern that all experts used to identify crater rims; in cases of rimless craters, especially in the WAC-based highlands counts, crater "rims" were marked as where the visible depression ends.

As is clear from Figs. 1 and 2, preservation state plays a role in the agreement between different persons in the crater identification and measurement. To explore this and quantify how it may contribute to scatter in crater identifications, we used a simple four-class system developed by Arthur *et al.* (1963) for the LPL Lunar Crater Catalog and used by Chapman since that time; it is summarized in Table 3. Since Chapman's identification technique includes classification of craters and he has been using the system the longest, his data were correlated with the reduced expert, volunteer, and matched expert-volunteer craters. From the NAC data, 14 expert craters, 57 volunteer craters, and 10 expert-volunteer matched craters did not have a corresponding match to Chapman's raw craters. From the WAC data, 193 (32%) ensemble craters $D \geq 7$ px could not be matched to Chapman's raw data. Robbins classified the missing craters in both instances. With these preservation state classifications, we conducted two separate tests to determine how crater identifications may depend on crater preservation.

## 6.1. What is the average preservation state as a function of crater diameter, and does this differ between experts and volunteers?

Fig. 10 shows the fraction of craters found per $2^{1/4}D$ multiplicative diameter bin per preservation state. The left column shows NAC data, and the right column shows WAC data. The general agreement between experts (top left) and volunteers (bottom left) shows that they generally found the same craters, though for craters $D \lesssim 50$ px, the most degraded craters (classes 3 and 4) comprised a smaller fraction of the volunteer dataset. This is consistent with persons who have more experience being able to consistently identify less-well-preserved craters.

For the NAC image, the data generally show that there are more large degraded craters and more small pristine-appearing craters. In Section 3.2, Antonenko suggested that craters in the ~30–80 px diameter range might be less preserved, overall, and so account for the lower number she found when ignoring heavily degraded craters (as was done when using the *ArcMap* interface). Fig. 10 shows that Antonenko was partly correct, and if the classes are reduced further such that class 1 is combined with 2, and 3 is combined with 4, then class 3+4 craters are the majority for $D \gtrsim 30$ px (for both experts and volunteers). There is no dominance by less pristine craters for $D \gtrsim 80$ px, however. It could be that the greater number of pixels, even for more degraded craters, when under-sampled on a computer screen makes their identification easier.

A hypothesis to explain why the experts disagreed more on the number of large craters in the NAC image was that, on a percentage basis, there were many more resurfaced craters at larger diameters. These data indicate that, to first order, this hypothesis is upheld. That is not to minimize the role of small numbers in the scatter of the crater counts, but these data show that this may be a contributing factor.

A hypothesis proposed in section 3.3 to explain the volunteer deficit of craters in the $D \sim 25$–40 px range relative to experts was that craters are more poorly preserved in this range. The data do not support this hypothesis: ~30 px is in the middle of this range (logarithmically) and is the cross-over point for where fresh craters are as numerous as degraded craters, meaning that in the $D \sim 25$–40 px range, the total number of well versus poorly preserved craters are comparable. Ergo, the hypothesis cannot be supported by the data and is at best ambiguous, and at worst it is rejected. A remaining possibility is psychological: Volunteers may be more attuned to identifying craters near the minimum size ($D = 18$ px) and near the maximum size ($D \sim 100$ px), but the intermediate diameters may be less noticeable to a layperson. A way to test this would be to have finer gradations in zoom level for each image rather than the 1×1, 3×3, and 9×9 that we currently use, and this may be addressed in future work (*i.e.*, a 1.5×1.5 and 2×2 zoom would put craters in the $D \sim 25$–40 px range near the minimum diameter).

WAC data are different. The mare data show similar numbers of all preservation classes across most diameters, except for $D \sim 10$–20 px where there are significantly more Class 1 and 2 craters than 3 and 4. The highlands data show that degraded craters dominate at smaller diameters (~60–80% vs. ~20–40%), and they are roughly equivalent for $D > 100$ px, though beyond this the error bars are too large to derive meaningful differences. This is likely not a reflection of the actual crater population but rather the ability to recognize craters. Large, degraded craters emplaced on terrain that has been reworked many times have few distinguishing features that permit identification and hence are less likely to be found. Ergo, we would expect a greater proportion of large, fresh craters on the lunar highlands to be found than the most degraded ones.

## 6.2. Does the scatter in crater identification and measurement depend on preservation state?

In Section 5, we discussed the scatter in individual craters as a function of crater diameter for experts, volunteers, and on mare versus highlands terrain, and this was illustrated in Fig. 8. Here, we expand this study, separating the craters by preservation state, and we illustrate the results in Fig. 11. Understanding how preservation state may or may not play a role in how well individuals agree upon a crater marking may allow us to work the problem backwards in the future: If a crater has little scatter in its location and diameter from volunteers, then it might be considered morphologically pristine, and vice versa, allowing us to automatically assign a preservation state.

However, the NAC results do not show this is feasible with volunteers. The bottom-left charts in Fig. 11 show no statistically significant difference in either crater location or diameter between the different preservation classes. This is in contrast with the top-left chart in Fig. 11 which show that there is a dependence for experts. Morphologically pristine craters have less than half the scatter in their locations and diameters amongst experts than those craters that are morphologically degraded; the NAC data also show that the relative scatter decreases with increasing diameter for pristine craters but remains fairly constant for degraded craters. In contrast, the WAC data show neither significant differences in highlands nor mare in the scatter in crater location and diameters when separated into preservation states. While this was expected based on Figs. 1 and 2, quantifying these data are nonetheless important. Additionally, the difference between experts and volunteers is important and somewhat counter-intuitive; this may be the study of future work and be a manifestation of how trained persons view craters versus minimally trained persons.

## 7. Results: Artifacts Near the Minimum Diameter

Throughout the preceding sections, we have identified several artifacts near minimum diameters. We discuss them in more detail here.

### 7.1. Are there issues with the clustering code at small diameters?

We point out in Sections 3.2, 3.3, and 5 that there are some artifacts near the minimum diameter of 18 px imposed in the NAC study. The most significant test of this was to cluster all expert data, cluster all expert data $D \geq 18$ px ("pre-clipped"), and compare the results. If there is no artifact, then the data will appear identical for $D \geq 18$ px. They did not. The data were identical (to within $<0.01\sigma$) for $D > 21.5$ px. For smaller diameters, the cluster results from the pre-clipped data had more craters per diameter than the other set, rising to $0.9\sigma$ different at $D = 18.7$ px at which point the number of clustered pre-clipped craters fell drastically until the value at $D = 18$ px was again nearly identical to the other set.

The reason for this artifact is explained by the following thought experiment: A crater with a "true" diameter of 18.1 px is identified by all experts. Due to the scatter between them, the experts measure the crater to have a diameter between 16.5 and 20 px, and 6 of the experts measure it as $\geq 18$ px. Considering all the data and clustering the markings together, a diameter close to 18.1 px is found. But, in the pre-clipped scenario, with only the $D \geq 18$ px markings, $N \geq N_{\text{threshold}}$ so the crater will be included in the final population, but the average of those 6 experts would be a crater closer to, *e.g.*, 19 px. From this, we conclude that an artificial small-diameter clip, when using any clustering code to group results together, will affect the results, aliasing crater diameters larger than they truly are. This means that for such endeavors, the true cutoff of data that can be believed will be larger than the minimum. In this study, that threshold was found to be 3.5 px above the minimum.

## 7.2. Does the Moon Mappers interface's strict 18-px cut-off affect results?

We establish in the preceding sub-section that the clustering code introduces artifacts near the minimum diameter. However, the nature of the Moon Mappers interface not permitting smaller crater measurements along with human psychology combine to create an even larger aliasing effect near the minimum diameter: People *want* to mark a crater if they have already started. The interface has a person draw a crater, and if it is <18 px across, the crater is red; it only turns green when the user makes it ≥18 px in diameter. For craters that are on that boundary of 18 px, one becomes psychologically invested in marking that crater if they have already spent the time clicking and dragging and will push it just large enough to count – and depending upon a person's eyesight, distance from their monitor, screen resolution, and manual dexterity, this may be up to several pixels. Both Antonenko and Robbins – despite being fully aware of this effect – found themselves doing it occasionally when marking craters. The effect is seen in the inset of Fig. 7 top-left panel. It shows that for craters $D > 22$ px, the experts match the clustered volunteer results well. But, below this, the number of craters per diameter rises much faster for the volunteer data, surpassing the $1.5\sigma$ level at $D \approx 19.5$ px, at which point it, like the pre-clipped expert data discussed in section 7.1, flattens with very few craters having a clustered diameter in the 18–19 px range. This is a confounding effect on top of the purely numerical one discussed above, and it emphasizes that even the raw data, when using an interface such as this, is not reliable for craters within ≈4 px of the minimum diameter.

## 7.3. Do integer diameters affect results?

All non-*ArcMap* interfaces had at least some rounding effects with crater diameters. With JMARS, every diameter was an integer value (in meters) for our NAC identifications, but this limitation was fixed by the time Kirchoff made her WAC identifications (and the artifact did not negatively affect the >100s m diameter craters visible in WAC for Antonenko's data). With Chapman's POINTS tool, many were near integers, and this was also the case with Moon Mappers. This means that at small diameters, crater CSFDs will show a step-like distribution (Fig. 7, top-left), though this can be smoothed if diameter bins are set large enough (Fig. 7, bottom). This affected our K-S tests (section 5): When we compared the results for different minimum crater diameter cut-offs between different expert and the volunteer CSFDs, we found that agreement improves (more comparisons designated "same" are found) as the minimum diameter is increased. This can be explained easily because the K-S test looks at the maximum deviation between two ranked normalized lists of data, and since the most craters are at small diameters, the largest differences will be found between the smallest craters from JMARS versus other interfaces.

## 7.4. How did each expert measure and ensure completeness?

Each expert was asked to provide NAC image crater counts for all craters $D \geq 18$ px and WAC counts to diameters "you are comfortable identifying." In every case for the former, each person said they identified craters several pixels smaller than the 18-px cut-off to assure completeness (the diameter to which we estimate that all craters were identified), and other than clustering and rounding artifacts, this was found to be an accurate method. We conclude from this that, if one is trying to ensure completeness to a certain minimum diameter $D_{min}$ (*i.e.*, they have included all craters $\geq D_{min}$), an individual should actually attempt to be complete to diameters a ~few px smaller than $D_{min}$. Conversely, if they think they are complete to a certain diameter, it is likely they have missed some craters, and $D_{min}$ is a few pixels larger.

The WAC experiment represented a very different case, and each person had a slightly different technique to estimate the smallest diameter to which they thought all craters were included (completeness):

- Antonenko: Estimated based on the maximum value in a ~100-m size bin on an incremental SFD. For her *ArcGIS* data, this is 400 m ($\gtrsim 7$ px), and for her JMARS data, this is 300 m ($\gtrsim 6$ px).

- Chapman: *A priori* estimate of ~12 px could be achieved, so measured down to 9–10 px ($\approx 600$ m) to ensure this.

- Fassett:  General "comfort" through experience, estimated at 700 m ($\gtrsim$11 px).

- Herrick:  8–10 px, visually estimated *a priori*, and so craters were measured to ~8 px.  He then looked at where the CSFD diverged from a straight line (on log-log axes) which was estimated as 600 m ($\gtrsim$9–10 px) after counts were completed.

- Kirchoff:  Measured craters down to 5 pixels across and then examined a CSFD of her data and looked for where the CSFD diverged from a straight line (on log-log axes).  Estimated to be 350 m ($\gtrsim$5.5 px) though cautioned this may be misleading.

- Robbins:  Created an incremental SFD in $2^{1/8}D$ multiplicative intervals, looked for the diameter bin with the largest number of craters, and estimated completeness to be the diameter bin one larger than that.  Also did this for individual 1000×1000 px latitude×longitude bins.  Most of the image was "complete" to 6–7 px, but the maximum was 9 px.

- Singer:  Normally would map no smaller than $\approx$5 px, but felt features as small as $\approx$4 px could be identified in the mare region.  Thus, she counted craters this small in the initial mapping and checked the small-diameter roll off in the SFDs later to estimate a completeness of ~5–6 px.

- Zanetti:  Estimated completeness to be ~500 m ($\gtrsim$8–9 px) in the mare area but due to the significant jumbled terrain, placed a very conservative completeness estimate of ~1 km ($\gtrsim$15–16 px) in the highlands.  These were estimated by comfort level.

These estimates are illustrated in Fig. 7 as small arrows on the CSFDs in the WAC panels.  With the reduced dataset, we can compare these estimates and determine their accuracy relative to the ensemble.  Chapman's, Fassett's, and Robbins' data overall and Kirchoff's mare counts show they under-estimated completeness (they were complete to smaller diameters than predicted) by a few pixels.  Kirchoff's highlands, both Antonenko's ArcGIS and JMARS mare and JMARS highlands, Herrick's mare, and Singer's mare counts show they had a good estimate of their completeness. Singer's fewer craters by a factor of 2× in the highlands relative to the ensemble show that she consistently under-counted craters relative to the ensemble, though the relative population was complete to the diameters she estimated (K-S test showed they are the same population). Herrick's estimate of complete counts for $D \geq 9$ px is also an over-estimate, for his counts have a shallower slope that deviates from the rest for $D < 15$ px which is the start of his lack of complete data. The CSFD slope of Antonenko's *ArcGIS* highlands counts begins to shallow relative to the other data for $D < 10$ px instead of her estimated ≈7 px completeness.

In the above bulleted list, it is clear that there are several different qualitative and quantitative methods employed by different individuals. With the criterion that the most conservative is the best estimate, Fassett's qualitative "gut feeling" was about as good as Robbins' complicated quantitative method. More concerning is the change in slopes near each person's minimum diameter, where Herrick's counts fall well below others' for ~6 px larger than his estimated completeness. Unfortunately, this is not the only kind of artifact observed: Both Kirchoff's and Robbins' highlands data instead show a steeper slope for $D < 10$ px and $D < 9$ px, respectively, relative to the ensemble results, indicating a possible aliasing effect to larger diameters. These effects were not dependent on the interface used to identify and measure craters.

Unfortunately, without counts in the region to even smaller diameters or a comparable dataset, it is not possible to state with absolute certainty the completeness of any one person's crater counts, and limitations in funding and time and the tediousness of identifying craters is a barrier to multiple experts wanting to duplicate another's counts. It is also probable that in any single person's crater counts anywhere near their estimated minimum diameter, any deviation from an expected crater population can easily be an artifact of being near that minimum, especially if it is fewer than 10 px in the image data that is being used. Extreme caution and a conservative approach are recommended near any minimum diameter estimate.

# 8. Results: Biases Causing the Primary Differences Among Experts and

## Between Experts and Volunteers

All tools except for the Moon Mappers interface are designed by professionals for professionals. Therefore, the largest *a priori* biases are likely to result from the MM interface versus the other interfaces. And, barring bugs (*i.e.*, JMARS rounding issue), this was the case based on previous Sections.

In the quest for a simple interface that provides the minimum learning curve, the MM marking tool lacks many of the image manipulation methods available to professionals. For example, the ability to zoom in and out – *i.e.* over- or under-sampling the image on-screen – is not available unless computer operating system- or internet browser-specific manipulations are enabled to permit one to zoom in. This could potentially result in less exact markings than desired, or more time might be needed to achieve the same desired precision. The interface also does not allow one to move a fraction of an image to the side, forcing one to sometimes identify only partial craters (however, the interface only allows one to identify craters where >50% is visible). While the image location is one type of limitation, a second is the way in which the image is displayed. As shown in Table A2, all experts in this project made use of image display modifications – be it individually or a combination of contrast, curves, gamma, greyscale inversion, brightness changes, or rotation – in order to enhance local topography differences to assist in crater identification. These tools are lacking in MM and so it is very much a "what you see is what you get" interface. Despite the inability to perform any image manipulation, the above analyses show that volunteers performed remarkably well as an ensemble in comparison with the experts as an ensemble. The interface was found to speed up counts by Antonenko while slow down counts by Robbins, but their results were comparable to their other techniques.

## 9. Discussion and Conclusions

This represents the first extensive study focused on understanding the difference in crater identification and measurement between numerous, independent crater experts using a variety of methods, and between those experts and lay volunteers. Throughout this work, we quantified several different comparisons and reached several conclusions:

1. We found anywhere from a ±10% to ±35% dispersion among experts in the number of craters found in different size bins, even when the numbers were many 10s or 100s of craters, and the crater diameters were well above the minimum diameter studied. This dispersion is dependent on the simplicity of the terrain with ~km-scale mare craters being the most consistent and the heavily modified lunar highlands the most varied.

2. The Moon Mappers online interface, despite its simplicity, yields results that are statistically as good as professionally developed interfaces for identifying craters, though artifacts are present within a few pixels of the minimum diameter (which could be eliminated by removing the minimum diameter or limiting studies to a few pixels larger than the minimum).

3. Except near the minimum diameter, volunteers are able to identify craters just as well as the experts (on average) when using the same interface (the Moon Mappers interface), resulting in not only a similar number of craters, but also a similar size distribution. This analysis was for the $N_{threshold}$ = 4 and 5 for experts (depending on number of datasets per image) and may vary if $N_{threshold}$ were different.

4. An *a priori* assumption that was nonetheless verified is that the lunar highlands crater counts are most prone to uncertainty – especially at ≤60° solar incidence – of the three types and scales of terrains studied, but the experts were conscious of this and placed caution on their exact crater counts. However, while future work may show this to be valid across a large number of sample terrains, it is technically valid only for this particular image set.

5. Experts and the individual volunteers differ significantly in that most experts found every crater that made it into the final catalog, but exponentially fewer volunteers found more craters that were in their final catalog.

6. Experts are more consistent among each other in crater measurement and

identification and are better at identifying degraded craters than volunteers: The scatter in location and diameter is around 5–10% and decreased with better preserved craters, but for volunteers, the scatter between individuals is approximately ±10% for location and ±20% for diameter and was mostly independent of crater preservation. There was no significant dependence on diameter. Ensemble crater diameters found by volunteers and experts matched 1:1 within their uncertainties.

7. Many artifacts occur near a minimum crater diameter, and experts are mixed in their ability to accurately assess their minimum completeness diameter. When the minimum diameter is well over 10 px, experts are able to consistently return complete counts by attempting to be complete to smaller diameters, but completeness at smaller than 10 px diameters returns artifacts that depend upon the person – *i.e.*, we found that some people identify more craters near that minimum, while others identify less. It is possible that different instructions (*i.e.*, "Data as accurate as possible for all $D \geq 4$ px craters are required for this work") may have resulted in more consistent small-diameter results, but the different artifacts found in different researchers' counts is still troublesome.

From these findings, we conclude that volunteers are approximately as good as experts in identifying craters, at least on terrain of "mixed" difficulty, so long as enough volunteers examine the image to derive a robust result (*e.g.*, at least 15 persons view the image; where this value may reach a point of diminishing returns will be the subject of future work). This finding is also with the caveats that (a) one should be interested in using the craters as an overall population as opposed to needing highly accurate results on just an individual or a few individual craters, and (b) they are willing to use craters a few pixels larger than the minimum identified by volunteers. From this work, we can also strongly recommend that researchers be cautious when using craters <10 px in diameter, especially in using small deviations from an expected function to conclude secondary crater contamination or crater erasure.

The variation in the number of craters identified by the experts on the same terrain has implications for interpretations that make assumptions about crater detectability. For example, Richardson (2009) modeled crater saturation, but inherent in that model is a detectability threshold to decide when a crater is sufficiently degraded or overprinted to be uncountable. Our data show that this threshold varies among experts. This factor is relevant to the longstanding debate about whether the heavily cratered terrains on the Moon and other planetary bodies are in saturation (Gault, 1970; Woronow, 1977, 1985; Hartmann, 1984; Squyres *et al.*, 1997; Richardson, 2009). While this is one example, the broader implication is that this variability would affect chronological information derived from complex surfaces, especially on more degraded terrain (*e.g.*, the lunar highlands).

With the caveat that secondary craters were not excluded, the extreme spread among experts in our example WAC highlands region at the 1-km-diameter point (where the lunar chronology is defined) is a factor of 4.5. Note, however, that all Poisson error bars in the counts overlap with at least one other expert's data except for Chapman's and Singer's ($D = 1$ km $= 15.5$ px, where the relative standard deviation is ±38% (Fig. 4)). While this work was not meant to derive model ages, we can look at the implications of it as applied to that process. The crater densities equate to a crater retention age of between 3.4±0.1 and 3.8±0.0 Gyr, while the ensemble is 3.7±0.0 Gyr (under the Hartmann (1999) system using the Ivanov (2001) chronology). The mare terrain, despite the overall agreement being better among experts, at $D = 1$ km has a density variation of 2.6. Due to the youth of this terrain, the resulting variation in ages is wider, from 1.3±0.4 to 2.2±0.5 Gyr, while the ensemble is 1.7±0.4 Gyr (the relative standard deviation is ±16% at this diameter, and all experts' error bars overlap each other (Fig. 4)). To add to the list of conclusions:

8. Age uncertainties based on counting statistics almost always are artificially small because they neglect natural variations in a single analyst's threshold of detection and the even larger variations expected for other expert analysts. Small differences in crater density between units measured by the same analyst may be real, but absolute ages based on an analyst's measured crater densities must be regarded as being subject to uncertainties at least as large as a factor of ±20% and even larger (>±35%) for difficult terrains such as lunar highlands.

The difference in the NAC data at 150 m (the smallest diameter that is not saturated, ≈220 px) is a factor of 2.3 between the minimum and maximum expert data. These correspond to ages of 1.5±0.7 to 3.2±0.8 Gyr. However, the variation between volunteers' and experts' ensemble results at this diameter is only a factor of 1.01 relative to each other, corresponding to an age estimate of $2.71^{+0.89}_{-0.91} - 2.72^{+0.89}_{-0.91}$ Gyr (three significant figures only used to illustrate the degree of similarity). While the error bars are quite large (few craters were $D \geq 150$ m, $\geq 220$ px in the NAC study area), this still shows significant variation on the order of at least several hundred million years is possible between mappers; we also found density variation to be generally greater than the factor of 2 which is inherent in the chronology function itself. Two additional conclusions can be drawn from this exercise.

9. It is not appropriate to quote model crater ages to three or more significant figures (though several of the authors of this paper are guilty of doing so, and we do so in the one comparison case above to show the excellent agreement). We recommend only using two significant figures.

10. Volunteers appear to be able to provide an ensemble result for age modeling as good as experts – despite the inherent uncertainties discussed throughout Sections 3-8 – in light of the larger uncertainties inherent to most applications of crater populations (such as age modeling).

The analysis presented in this work relied on a clustering algorithm and setting an $N_{threshold}$ of the number of persons who must have identified a crater for it to be included in the "ensemble" catalog. For the experts, we set this at approximately half the number of datasets we had assembled. However, one could plausibly argue that $N_{threshold}$ should be set to a smaller number, such as 2, or a larger number. Changing the value will alter the number of craters in the final catalog, and raising or lowering it would represent two different philosophies: If one requires fewer analysts to have found the crater, then this could be interpreted as it being unlikely that two people would independently find the same "non-crater" and it is more likely that others simply missed it. At the other extreme, requiring more people to have found the feature for it to count takes a more conservative approach where one would want to be much more certain the feature is truly a crater before adding it to a final catalog. Each has their own separate uses, and one should select an $N_{\text{threshold}}$ that is most appropriate for their application. In this work, we used $N_{\text{threshold}} \approx 0.5 N_{\text{experts}}$ so as to not weight any one analyst more than others and produce a catalog of craters that the "average" analyst should find. The number of craters we identified with this $N_{\text{threshold}}$ value is almost certainly less than the objective reality in that there will be craters that are physically present that most persons will not find, and perhaps no analyst could find. But, we think this represents a reasonable consensus of the crater population for these surfaces.

This study presents the most vetted catalog of craters on two targeted regions of the lunar surface with a range of morphology. This kind of data product is useful as a training set for individuals learning how to identify craters, calibrating themselves, and for training automated crater detection codes. As a resource for those communities, we have made these data and images available as a supplemental material with this paper, and we encourage the reader to contact the first author for more details.

While this work has focused on the Moon, we expect that its results are applicable to at least other airless bodies and likely to all planetary surfaces in general because crater identification and measurement requires similar skills from one surface to another. The main consideration for applying these results to other worlds would be to account for the role of active geological processes that affect craters in various ways and whether there are other endogenic processes (*e.g.*, collapse pits) that mimic the appearance of impact craters. Additionally, different gravity fields and extreme local slopes can alter crater morphology. For example, Mars has fluvial and aeolian processes to consider when determining whether a feature is a degraded crater, and asteroid (4) Vesta has numerous craters emplaced on steep slopes that result in highly asymmetric features that an untrained person may not recognize as an impact crater. Two

Venusian crater catalogs constructed in the 1990s (Schaber *et al*., 1992; Herrick and Phillips, 1994) had differences based on what each group considered an impact crater versus a volcanic structure, a difficulty enhanced by atmospheric fragmentation clustered smaller impact craters together, forming irregular features that were easily mistaken for irregular volcanic calderas. Overall, however, the lessons learned from this work on lunar craters should be applicable to most planetary bodies, and the same cautions and considerations should be applied.

# Appendix A: Details of Crater Identification and Measurement Methods

Most of the researchers in this study used different interfaces and measurement techniques. They are described briefly in Section 2.2 and in depth in this Appendix.

## A.1. ArcGIS

ArcGIS software, produced by ESRI, is a popular suite in the geology field and contains numerous tools and add-on software that are useful in analyzing geographic datasets and annotating them. The majority of experts in this study used ArcGIS software, though four distinct methods were employed within the *ArcMap* software application.

## A.1.1. Rim-Tracing

Robbins was one of two researchers to use the native tools within *ArcMap*. His technique is more thoroughly described in Robbins and Hynek (2012), but briefly, each image was imported into an empty *ArcMap* file without any ancillary projection files (*i.e.*, the PGW file – PNG World file – was removed). This forced *ArcMap* to treat the image as being in pixel space. He used *ArcMap*'s Editing tools in "streaming" mode to lay down a vertex every 2.5 pixels – this was chosen because it would be about 20 points along the rim of the smallest crater (on NAC), and it allowed him to not be pixel-perfect in vertex deposition. The vertex spacing was decreased to permit smaller crater identification on the WAC image. He then traced every crater rim using a Wacom tablet and pen as an input device. These were saved to a GIS shapefile and, when finished, exported as a text file in units of pixels. He then used custom software to fit both circles and ellipses to each crater rim.

Robbins was the fastest crater identifier in this study, probably for two reasons. First, he uses a pen/tablet input device, allowing a more natural drawing capability; he was one of only two researchers in this study to use a pen (Singer was the other, and her identifications were second-fastest). Second, he traces crater rims which is simply an act of drawing a circle with the pen and pressing the "F2" or a software-assigned button on the tablet to close the circle. He can then immediately move onto the next crater with less than one second pause between. In this way, he does not have to worry about how a three-point fit looks nor manually click three or more points along the rim, it's simply draw-and-click, draw-and-click, etc.

## A.1.2. CraterTools

The CraterTools extension to *ArcMap* was released by Kneissl *et al*. (2011). It includes a three-point tool that fits circles similar to other interfaces discussed in this Appendix: The researcher manually identifies three points along the rim and then CraterTools fits a circle and records that in a GIS shapefile.

Fassett's modification to this was to overlay a grid on top of the image at ≈70% the minimum diameter desired. Any crater that is smaller than a single grid cell is ignored.

## A.1.3. Crater Helper Tools

This USGS extension to *ArcMap* was used by Antonenko and Herrick to manually identify craters (Herrick used this for the NAC exercise and CraterTools for WAC). Craters are identified by clicking three points on the rim, which uniquely identifies a circle. In all cases, craters were fit with a circle, and the center point and diameter of the circle are then provided by the software.

## A.1.4. Chord-Drawing

Singer also used the native tools in *ArcMap* to map craters with an additional extension to calculate spherical lengths (Tools for Graphics and Shapes (Jenness, 2011)). Crater diameters were measured by drawing a straight line from one rim to another (one point on each side), going through the center of the crater, with an attempt at consistency about orientation of the line with respect to the lighting geometry (bisecting the shadow in most cases). This will likely give the most accurate estimate of the diameter. There may be some advantages and disadvantages of this technique as compared to those that fit circles to a number of points on the rim – it might seem that precision could be sacrificed for speed. From a visual assessment of Figs. 1 and 2, however, there is no obvious trend for craters mapped this way to be consistently off from the overall average. Large discrepancies between mapped diameters may be due to individual discretion about the location of the rim, especially for degraded craters. One advantage may be that measurements were made relatively consistently with respect to lighting geometry which may or may not be the case when marking three points on a rim via other techniques. Singer also made use of a tablet and pen input device.

## A.2. JMARS

JMARS is a tool produced and maintained by Arizona State University that many use as a free alternative to ArcGIS software for displaying planetary datasets. It has a built-in crater-measuring tool. Craters are identified visually and the "Three Point Mode" of the tool was used by Kirchoff. Three points along the identified rim are manually selected and then the circle-fit calculation is performed by JMARS. Antonenko used the circle-drawing tool contained within JMARS which is a similar method to the Moon Mappers interface. Craters are digitally recorded by JMARS with a center latitude and longitude in degrees and diameter in meters. A minimum size can be set through the use of a smaller-than-minimum template (the "Add Mode" in the JMARS crater counting tool) to determine which craters are equal to or larger than this diameter.

A limit of the version of JMARS prior to early April 2013 was that crater diameters were rounded to the nearest integer measurement (meter) when using the 3-point and circle-drawing tools (sub-meter measurements can be made with the "add" tool), although locations are not rounded. This caused aliasing in the small-diameter range in later analysis. The maintainers of JMARS were aware of this issue and fixed it in early April 2013 – too late for the NAC data, but in time for Kirchoff's WAC data.

## A.3. SAOImage DS9

The DS9 interface was used by Chapman in conjunction with Peter Tamblyn's crater measurement program (written in Perl and running on Windows XP, with subroutines from the POINTS software developed at Cornell by J. Joseph and P. Thomas in the late 1990s). The crater measurement loop works as follows: The measurer positions the tip of the cursor arrow at a point on the rim of a crater and clicks it, producing an X-shaped mark; this is done 2, 4, 6, or more times (3 and 5 times don't work). After marking the rim, the measurer moves the cursor to a small window, clicks within it to make it active, types "f" and a number from 1 to 4 (representing the morphology classification estimated by the measurer), hits carriage return, then "l". At this time, the X's in the main image are replaced with the fitted ellipse. The $\{x,y\}$ position and semi-major and –minor axes are written to a data file. If an error is perceived, there are buttons in the "craters" window that can change the shape, size, and position of the fitted ellipse (very rarely used); alternatively, the crater can be deleted and a new set of rim measurements made from scratch (such deletions cause gaps in the crater numbering). With the cursor positioned on the rim of the next crater, a click begins the process for the next crater.

There is a coding issue when a crater is being measured that overlaps, or lies within, a previously measured crater. There is an "overlay on/off" button that changes the measurement mode so that it works; the cursor has a different appearance, but otherwise is the same from the measurer's perspective. The DS9 viewer window can be zoomed in or out by a factor of two in scale, although measurements can be made only in zoom = 1, 2, 4, etc. (not ½, ¼, etc.). At larger zoom values, the position of the measurement window can be adjusted to the correct part of the whole image using a small window that shows the viewed portion in a blue rectangle. With the right button pressed, dragging the cursor around the image can change the brightness and contrast of the image, which especially affects the visibility of shallow craters.

## A.4. Moon Mappers Interface

The Moon Mappers interface was used exclusively by volunteers. For comparison purposes, both Antonenko and Robbins (science co-leads on Moon Mappers) also used this interface to determine if there were significant differences between it and their preferred method.

This is an internet-based interface. After logging in at cosmoquest.org, volunteers select the "Moon Mappers" application and can choose a "Simply Craters" or "Man vs. Machine" interface. In the latter, images have pre-marked craters overlaid from an automated crater detection code; craters identified within that interface were excluded from this analysis because this approach did not match the process used by experts and would have added complexity to the comparison. For both interfaces, images are prepared by a sub-dividing code. First, the NAC image is scaled to 100%, 33%, and 11% size (1×1, 3×3, and 9×9 pixel binning). Each scaled NAC is then divided into 450×450-pixel sub-images with a minimum of 30 pixels overlap on the edges. This is meant to ensure that no portion of an image is missed and that they are viewed at multiple scales. For the portion of the M146959973L image used in this study, only 1×1 and 3×3 binning was applied because of the small width of the study region.

Within the MM online interface, each sub-image is displayed to the volunteer one at a time in quasi-random order. The interface has a basic crater-marking tool that allows the user to draw a circle by clicking at the crater center and dragging outwards. The circle is red until it reaches the minimum diameter of 18 pixels, and then it turns green. If the mouse button is released while the circle is red, it does not save; green, it does. The volunteer can use another tool to reposition the final circle and resize it. Users can mark craters as small as 18 pixels across or as large as 450 pixels across. When they are satisfied they have annotated all $D \geq 18$ px craters on an image, a "Done Working" button is pressed and the craters are saved to an online database.

## Appendix B:  Details of Researcher Experience

One of the primary purposes of this study is to identify variation in crater identification among experts.  This Appendix provides brief biographical notes about each researcher that focuses on their crater identification experience.  Researchers are listed alphabetically by surname, not in author order.

**Antonenko** has over 20 years experience in identifying and measuring craters on the Moon, using a variety of data sets, resolutions, and lightning conditions to determine the sub-surface stratigraphy revealed by impact craters.  Other work by Antonenko has also involved studying craters on Venus, Ganymede, and the Earth.  Antonenko, along with Robbins, is Science co-Lead for Moon Mappers.

**Chapman** began measuring craters within weeks of graduating high school in 1962, and he has had experience measuring craters on ten bodies (Moon, Mars, Gaspra, Ida, Europa, Ganymede, Callisto, Mathilde, Eros, Mercury), in recent decades as a member of numerous spacecraft science teams.  With a career in this work spanning over five decades, he is the most senior researcher involved in this work and has used the broadest range of datasets (quality, surface, medium) and techniques.

**Fassett** has eight years (since 2006) experience measuring craters.  Along with Seth Kadish, he constructed a catalog of ~15,000 craters with $D \geq 20$ km on Mercury and the Moon. A counter-point to this experience is that he almost never works with craters smaller than 100 m, as emphasized in Table A1 where he considers the NAC image both difficult and of little chronologic utility because of its saturated state.

**Herrick** has been involved in impact cratering studies for ~20 years.  He participated in creating the first global catalog of Venusian craters using radar images from the *Magellan* mission, and he has catalogued craters on other planetary bodies for various research tasks.

**Kirchoff** has seven years (since 2007) experience identifying craters on various surfaces including the Moon, Mars, and icy satellites of Jupiter and Saturn. This work has helped determine the geologic and bombardment histories of these bodies while also generating global crater databases for several icy satellites of Saturn.

**Robbins** has seven years (since 2007) experience identifying craters on inner solar system bodies with most work focusing on Mars. He manually constructed a ~640,000 Martian crater database, published in 2012 (Robbins and Hynek, 2012) and has worked on similar analyses on lunar craters. Along with Antonenko, Robbins is Science co-Lead for Moon Mappers.

**Singer** has seven years (since 2007) experience mapping and measuring various geologic features (circular or otherwise) on the icy satellites in the outer solar system. This includes craters on various bodies (Europa, Ganymede, Enceladus, and Triton) as well as other geologic features on Europa (sub-circular chaos, pits, uplifts) and Iapetus (lineaments and long-runout landslides).

**Zanetti** has six years (since 2008) experience studying the formation and degradation of impact craters on the Moon, Mars, and Earth and has counted craters on volcanic structures and crater ejecta blankets using high resolution LROC NAC images of the Moon (Zanetti *et al.*, 2013) and HiRISE images of young surfaces on Mars.

## Appendix C: Details of Crater Clustering Code

In this Appendix, we detail how our clustering code works to automatically group features into single craters. We started from the basic two-dimensional DBSCAN algorithm of Ester *et al.* (1996). The original DBSCAN code was developed to (a) require a minimum amount of *a priori* knowledge (represented by a minimum number of input parameters), (b) identify clusters of arbitrary shape, and (c) scale well with large datasets. This algorithm was chosen over other potential codes (such as expectation-maximization (EM) clustering) because of the nature of our data: The density of crater markings varies as a function of feature size (with larger features having more absolute spread) and sets of markings have a small sample size (≈15 volunteers view each image) that are not necessarily normally distributed. The original DBSCAN algorithm hinges on the idea of "reachability," where one datum is considered "reachable" by another if its location is within a distance $\varepsilon$ specified *a priori* by the user. If enough points are reachable by each other, they are considered to be members of a cluster. The minimum number of points ($N_{\text{threshold}}$) needed to be considered a cluster is also pre-defined by the user.

The advantage of DBSCAN is that the algorithm is straightforward, easy to implement, and it is accurate enough for the next step – the science analysis. However, a drawback is the time complexity (how long the code takes to run) which is $\mathcal{O}(N^3)$. First, iterating through each point in the dataset in order to find matches takes $\mathcal{O}(N)$. The search to find reachable points by the point in question is another $\mathcal{O}(N)$. If a reachable point is found, since other points not previously reachable may now be reachable by the larger cluster, the code must again search through every point which adds another $\mathcal{O}(N)$ that brings the total time complexity to $\mathcal{O}(N^3)$. Due to the large datasets that we currently have and which we anticipate generating as an outcome of our volunteers' efforts, we have implemented an indexing structure that decreases the time complexity.

DBSCAN, as originally formulated, is a two-dimensional algorithm that works only with $\{x, y\}$ location data and compares this to the distance threshold $\varepsilon$. In our work, the reality of planetary surface features requires diameter to be considered, as well. This allows us to correctly address situations such as when a small crater is superposed on a larger crater. A traditional DBSCAN code would merge these two features together because they are within each others' $\varepsilon$ reachability, but their diameters clearly distinguish them as separate features.

In 2D DBSCAN codes, two points $p_i$ and $p_j$ are considered part of the same cluster as follows:

```
if(distance(pi,pj)<ε1) then pi and pj are clustered
```

To take into consideration both distance and diameter, we modified the if() statement such that two points $p_i$ and $p_j$ are considered part of the same cluster when the following is true:

```
if( (distance(pi,pj)<ε1*(Di+Dj)/2) and (|Di-Dj|<ε2*min(Di,Dj)) )
```

As before, the first criterion compares the distance between points $p_i$ and $p_j$, and the user-defined $\varepsilon_1$ parameter, but in our implementation, this parameter is scaled by the average diameters of the two features in question. This scaling accounts for scatter in feature location that varies with feature size rather than being an absolute measured value (*e.g.*, scatter can be ±5 pixels for a 50-pixel-diameter feature and ±10 pixels of a 100-pixel-diameter feature rather than ±5 pixels for both). The second criterion compares the feature diameters – our third dimension – where the difference of the two diameters is compared with the smaller feature's diameter multiplied by a second reachability parameter, $\varepsilon_2$. This second criterion needs to be a relative, non-dimensional term. Otherwise, $\varepsilon_2$ would need to be different for any given size crater (*e.g.*, a 3-pixel diameter difference from a ~5-pixel-diameter crater would not be considered a match, but a 3-pixel diameter difference between ~200-pixel-diameter craters is negligible and should be considered a match).

We determined the optimum reachability parameters by examining a large dataset from an automated crater detection output for a related project. This set contained approx. 150,000 crater markings that reduced to 10,104 clustered craters and the minimum number of features in each cluster was $N_{\text{threshold}} \geq 5$. To determine the optimum parameters, first, the physical distance, in pixel space, was determined between one crater and every other crater in the dataset. For each crater pair considered, the distance was normalized by the average diameter of the two craters being compared. A histogram of the normalized distances was then plotted. This showed a peak at values very close to 0 that decreased to a minimum at 0.35–0.5, and it then rose to encompass more craters. This minimum was robust across different datasets from our projects – including both the expert and volunteer data in this work. We interpret this as the large number of craters that have distances very close together are actual matches for the same crater, and that the histogram's minimum is the maximum distance craters can be separated before they are no longer likely to be in the same cluster. Ergo, this minimum of 0.5 was chosen as our reachability for distance, $\varepsilon_1$.

Of the pairs for which $\varepsilon_1 < 0.5$, the non-dimensionalized diameter reachability parameter $\varepsilon_2$ was calculated:

$$\varepsilon_2 > |D_i - D_j| / \min(D_i, D_j) \tag{1}$$

After much testing of this and numerous other potential non-dimensionalizations, this parameter was found to robustly determine whether two craters that were within the $\varepsilon_1$ threshold for location should be members of the cluster if, conveniently, $\varepsilon_2 < 0.5$, as well. Since division is often more time-consuming than multiplication, this leads to the second part of our earlier pseudocode: ($|D_i-D_j| < \varepsilon_2*\min(D_i,D_j)$). It should be noted that the value of $\varepsilon_2$ can vary up to ~1.0 without significantly affecting the results from our data.

The inputs to the clustering code are a tab-separated list of crater location $\{x, y\}$, diameters $D$, and confidence $c$. Confidence is a scale of 0–1 and used for MM volunteers to score how well each volunteer performed relative to expert markings (Robbins') on calibration images. For experts, we set $c = 1$. After completion, the data output for each cluster of markings are: (a) weighted mean $\overline{x}$ with standard deviation, $\delta x$; (b) weighted mean $\overline{y}$ with standard deviation, $\delta y$; (c) weighted mean $\overline{D}$ with standard deviation, $\delta D$; (d) number of points $N$ in that cluster; and (e) weighted mean of the confidence for the craters that went into that cluster, $\overline{c}$. All weights are based on the confidence $c$.

Acknowledgements: The authors thank two anonymous reviewers for their helpful feedback. Support for Robbins was through the Maryland Space Grant Consortium and the NASA Lunar Science Institute Central office. Support for Kirchoff and Chapman was through the Center for Lunar Origin and Evolution node of the NASA Lunar Science Institute. Support for Zanetti was through the McDonnell Center for Space Science, Lawrence Haskin Memorial Fellowship. Support for Lehan, Huang, and Gay was through support for CosmoQuest by NASA grants NNX09AD34G and NNX12AB92G, STScI under NASA contract NAS 5-26555, and funding through the *Lunar Reconnaissance Orbiter* education and public outreach office.

# References

Arthur, D.W.G., Agnieray, A.P., Horvath, R.A., Wood, C.A., Chapman, C.R., 1963. The system of lunar craters, quadrant I. Comm. Lunar Planet. Lab. 2, #30.

Crater Analysis Techniques Working Group, 1979. Standard techniques for presentation and analysis of crater size-frequency data. Icarus 37, 467–474. doi:10.1016/0019-1035(79)90009-5.

Ester, M., Kriegel, H.-P., Sander, J. and X. Xu, 1996. A density-based algorithm for discovering clusters in large spatial databases with noise. Proc. 2nd Int. Conf. on Knowledge Disc. and Data Mining, AAAI Press. 226–231. ISBN 1-57735-004-9.

Gault, D.E., 1970. Saturation and equilibrium conditions for impact cratering on the lunar surface: Criteria and implications. Radio Sci., 5, 273-291. doi: 10.1029/RS005i002p00273.

Greeley, R., and D.E. Gault, 1970. Precision size-frequency distributions of craters for 12 selected areas of the lunar surface. Moon, 2, 10-77. doi: 10.1007/BF00561875.

Hartmann, W.K., 1984. Does crater "Saturation Equilibrium" occur in the Solar System? Icarus, 60, 56-74. doi: 10.1016/0019-1035(84)90138-6.

Hartmann, W.K., 1999. Martian cratering VI: Crater count isochrons and evidence for recent volcanism from Mars Global Surveyor. Meteorit. Planet. Sci., 34, 167-177.

Herrick, R.R., and R.J. Phillips, 1994. Implications of a global survey of Venusian impact craters. Icarus, 111, 387-416. doi: 10.1006/icar.1994.1152.

Hiesinger, H., van der Bogert, C.H., Pasckert, J.H., Funcke, L., Giacomini, L., Ostrach, L.R., and M.S. Robinson, 2012. How old are young lunar craters? J. Geophys. Res., 117, CiteID E00H10. doi: 10.1029/2011JE003935.

Ivanov, B.A., 2001. Mars/Moon cratering rate ratio estimates. Chronology and Evol. of Mars, 96, 87-104.

Jenness, J., 2011. Tools for graphics and shapes: Extension for ArcGIS. Jenness Enterprises.

URL: http://www.jennessent.com/arcgis/shapes_graphics.htm

Kirchoff, M., Sherman, K., and C.R. Chapman, 2011.  Examining lunar impactor population and evolution: Additional results from crater distributions on diverse terrains.  DPS.

Kneissl, T., van Gasselt, S., and G. Neukum, 2011.  Map-projection-independent crater size-frequency determination in GIS environments—New software tool for ArcGIS.  Planet. Space Sci., 59, 1243-1254.  doi: 10.1011/j.pss.2010.03.015.

McGill, G.E., 1977.  Craters as "fossils:" The remote dating of planetary surface materials.  GSA Bull., 88, 1102-1110.  doi: 10.1130/0016-7606(1977)88<1102:CAFTRD>2.0.CO;2.

Nava, R.A., 2011.  Crater Helper Tools for ArcGIS 10.  MRCTR GIS Lab, Astrogeology Science Team, U.S. Geological Survey, Flagstaff, AZ.  URL: http://astrogeology.usgs.gov/facilities/mrctr/gis-tools

Ostrach, L.R., Robinson, M.S., Denevi, B.W., and P.C. Thomas, 2011.  Effects of incidence angle on crater counting.  Planetary Crater Conf., 2, Abstract #1107.

Richardson, J.E., 2009.  Crater saturation and equilibrium: A new model looks at an old problem.  Icarus, 204, 697-715.  doi: 10.1016/j.icarus.2009.07.029.

Robbins, S.J., Antonenko, I., Lehan, C., Moore, J., Huang, D., and P.L. Gay, 2012.  CosmoQuest MoonMappers: Cataloging the Moon.  Lunar Science Forum, 5, Abstract #602.

Robbins, S.J., and B.M. Hynek, 2012.  A new global database of Mars impact craters ≥1 km: 1. Database creation, properties, and parameters.  J. Geophys. Res., 117, CiteID E05004.  doi: 10.1029/2011JE003966.

Schaber, G.G., Strom, G.H., Moore, H.J.,, Soderblom, L.A., Kirk, R.L., Chadwick, D.J., Dawson, D.D., Gaddis, L.R., Boyce, J.M., and J. Russell, 1992.  Geology and distribution of impact craters on Venus: What are they telling us?  J. Geophys. Res., 97:E8, 13,257-13,301.  doi: 10.1029/92JE01246.

Shoemaker, E.M., 1965.  Preliminary analysis of the fine structure of the lunar surface in Mare Cognitum.  In: Heiss, W.N., Menzel, D.R., O'Keefe, J.A. (Eds.), The Nature of the Lunar Surface. Johns Hopkins Press, Baltimore, pp. 23–77 (JPL Tech. Report no. 32-700).

Squyres, S.W., Howell, C., and M.C. Liu, 1997. Investigation of crater "saturation" using spatial statistics. Icarus, 125, 67-82. doi: 10.1006/icar.1996.5560.

Wilcox, B.B., Robinson, M.S., Thomas, P.C., and B.R. Hawke, 2005. Constraints on the depth and variability of the lunar regolith. Meteorit. Planet. Sci., 40:5, 695-710. doi: 10.1111/j.1945-5100.2005.tb00974.x.

Woronow, A., 1977. Crater saturation and equilibrium: A Monte Carlo simulation. J. Geophys. Res., 82, 17, 2447-2456. doi: 10.1029/JB0882i017p02447.

Woronow, A., 1985. A Monte Carlo study of parameters affecting computer simulations of crater saturation density. Proc. Lunar Planet. Sci. Conf., 15, in J. Geophys. Res., 90, C817-C824. doi: 10.1029/JB090iS02p0C817.

Zanetti, M., Jolliff, B., van der Bogert, C.H., and H. Hiesinger, 2013. New determination of crater size-frequency distribution variation on continuous ejecta deposits: Results from Artistarchus crater. Lunar & Planet. Sci. Conf., 44, Abstract #1842.

Table 1: Summary of the numbers of craters found in the NAC image larger than or equal to different pixel cut-offs by the different researchers in the different interfaces. "Std. Dev." are standard deviations ( $\sqrt{N^{-1}\sum(x_i-\mu)^2}$ ) and "Rel Std. Dev." is the standard deviation normalized by the mean ( $\mu^{-1}\sqrt{N^{-1}\sum(x_i-\mu)^2}$ ). Also shown are different methods of determining the "best" number of craters in the image. Boxed values for "$N \geq$ # people" correspond to the best matches with the mean, and the bolded row is the row used for all analyses in the manuscript. Caution is advised when examining numbers of craters $D < 25$ pixels due to aliasing (see Section 7) which is why they are shown in grey.

| | | $D \geq$ 100 px | $D \geq 50$ px | $D \geq 30$ px | $D \geq 25$ px | $D \geq 20$ px | $D \geq 18$ px |
|---|---|---|---|---|---|---|---|
| Antonenko | Crater Helper Tools | 33 | 90 | 208 | 333 | 532 | 636 |
| Antonenko | JMARS | 34 | 119 | 254 | 387 | 616 | 945 |
| Antonenko | MoonMappers | 41 | 123 | 255 | 341 | 527 | 653 |
| Chapman | DS9 | 52 | 173 | 417 | 580 | 951 | 1197 |
| Fassett | CraterTools | 36 | 104 | 281 | 397 | 659 | 815 |
| Herrick | Crater Helper Tools | 46 | 137 | 327 | 475 | 704 | 809 |
| Kirchoff | JMARS | 26 | 92 | 253 | 360 | 570 | 831 |
| Robbins | ArcMap | 46 | 142 | 352 | 508 | 830 | 1060 |
| Robbins | MoonMappers | 47 | 151 | 333 | 483 | 784 | 991 |
| Singer | Geodesic Tools | 21 | 73 | 184 | 299 | 526 | 652 |
| Zanetti | CraterTools | 59 | 179 | 363 | 502 | 769 | 930 |
| **All Data** | **Mean:** | 40 | 126 | 293 | 424 | 679 | 865 |
| | **Median:** | 41 | 123 | 281 | 397 | 659 | 831 |
| | **Std. Dev.:** | 11 | 34 | 71 | 90 | 142 | 180 |
| | **Rel Std. Dev.:** | **28.2%** | **27.4%** | **24.2%** | **21.2%** | **20.9%** | **20.8%** |
| Experts (11 Interfaces) | $N \geq 6$ people | 35 | 122 | 276 | 408 | 662 | 836 |
| | $N \geq 5$ people | **41** | **135** | **295** | **431** | **700** | **889** |
| | $N \geq 4$ people | 47 | 148 | 322 | 459 | 742 | 943 |
| No Expert's Moon Mapper Data | **Mean:** | 39 | 123 | 293 | 427 | 684 | 875 |
| | **Median:** | 36 | 119 | 281 | 397 | 659 | 831 |
| | **Std. Dev.:** | 12 | 37 | 77 | 94 | 144 | 181 |
| | **Rel Std. Dev.:** | **31.5%** | **30.3%** | **26.3%** | **21.9%** | **21.1%** | **20.7%** |
| Experts (9 Interfaces) | $N \geq 6$ people | 32 | 102 | 249 | 376 | 618 | 780 |
| | $N \geq 5$ people | **37** | **122** | **280** | **419** | **675** | **850** |

| | | | | | | |
|---|---|---|---|---|---|---|
| $N \geq 4$ people | 41 | 132 | 303 | 444 | 718 | 905 |

Volunteers (Moon Mappers data as of April 4, 2013)

| | | | | | | |
|---|---|---|---|---|---|---|
| $N \geq 9$ people | 35 | 125 | 232 | 345 | 603 | 654 |
| $N \geq 8$ people | 38 | 134 | 248 | 365 | 661 | 730 |
| **$N \geq 7$ people** | **41** | **143** | **264** | **388** | **723** | **813** |
| $N \geq 6$ people | 43 | 153 | 288 | 418 | 788 | 916 |
| $N \geq 5$ people | 46 | 161 | 311 | 449 | 863 | 1043 |

Table 2: The same as Table 1 except for WAC data. Top section is from the mare region and bottom section is from the highlands region.

| **Mare Region** | | $D \geq$ 100 px | $D \geq 50$ px | $D \geq 20$ px | $D \geq 10$ px | $D \geq 5$ px |
|---|---|---|---|---|---|---|
| Antonenko | Crater Helper Tools | 0 | 3 | 9 | 92 | 496 |
| Antonenko | JMARS | 0 | 3 | 10 | 89 | 624 |
| Chapman | DS9 | 0 | 1 | 7 | 99 | N/A |
| Fassett | CraterTools | 0 | 2 | 8 | 111 | 657 |
| Herrick | Crater Helper Tools | 0 | 3 | 9 | 94 | 193 |
| Kirchoff | JMARS | 0 | 0 | 6 | 77 | 801 |
| Robbins | ArcMap | 2 | 4 | 10 | 117 | 699 |
| Singer | Geodesic Tools | 0 | 0 | 5 | 90 | 633 |
| Zanetti | CraterTools | 0 | 2 | 11 | 97 | 639 |
| | | | | | | |
| All Data | **Mean:** | 0 | 2 | 8 | 96 | 593 |
| | **Median:** | 0 | 2 | 9 | 94 | 636 |
| | **Std. Dev.:** | 1 | 1 | 2 | 12 | 182 |
| | **Rel Std. Dev.:** | **300.0%** | **72.7%** | **21.2%** | **12.3%** | **30.8%** |
| | | | | | | |
| Experts (9 Interfaces) | $N \geq 5$ people | | 2 | 8 | 93 | 570 |
| | **$N \geq 4$ people** | | **2** | **8** | **95** | **662** |
| | N $\geq$ 3 people | | 2 | 8 | 96 | 729 |

| **Highland Region** | | $D \geq$ 100 px | $D \geq 50$ px | $D \geq 20$ px | $D \geq 10$ px | $D \geq 5$ px |
|---|---|---|---|---|---|---|
| Antonenko | Crater Helper Tools | 17 | 28 | 111 | 210 | 369 |
| Antonenko | JMARS | 13 | 25 | 95 | 213 | 461 |
| Chapman | DS9 | 14 | 32 | 143 | 364 | |
| Fassett | CraterTools | 11 | 19 | 66 | 151 | 337 |
| Herrick | Crater Helper Tools | 15 | 22 | 72 | 122 | (None) |
| Kirchoff | JMARS | 13 | 20 | 68 | 191 | 503 |
| Robbins | ArcMap | 14 | 32 | 93 | 222 | 412 |
| Singer | Geodesic Tools | 9 | 14 | 32 | 170 | 170 |
| Zanetti | CraterTools | 5 | 12 | 55 | 148 | 336 |
| | | | | | | |
| All Data | **Mean:** | 12 | 23 | 82 | 199 | 370 |
| | **Median:** | 13 | 22 | 72 | 191 | 369 |
| | **Std. Dev.:** | 3 | 7 | 33 | 70 | 108 |
| | **Rel Std. Dev.:** | **27.3%** | **31.6%** | **40.2%** | **35.4%** | **29.3%** |

| Experts (9 Interfaces) | | | | | |
|---|---|---|---|---|---|
| $N \geq 5$ people | 12 | 20 | 68 | 160 | 300 |
| **$N \geq 4$ people** | **12** | **21** | **79** | **187** | **370** |
| $N \geq 3$ people | 13 | 22 | 90 | 221 | 432 |

Table 3:  Crater preservation state classification system.  Classes are at the top.

| | Class 1 | Class 2 | Class 3 | Class 4 |
|---|---|---|---|---|
| Rim | Sharp | Muted but still distinct | Wide, muted, some topographic expression | Barely distinguishable |
| Ejecta | At least some present | Possibly present | None | None |
| Walls | Fresh | Softened | Mantled-looking | Shallow or non-existent; mantled |
| Floor | Clear, or possibly flat with boulders and impact melt | Clear or have a few deposits | Not distinct from walls | Shallow, buried or mantled, not distinct from walls |
| Shape | Bowl / concave | Slightly shallower bowl | Shallower bowl | Barely discernible depression |

Table A1: Key data on crater identification for each expert for both the NAC and WAC images.

| Researcher | Time to Analyze NAC | Time to Analyze WAC | $N_{\geq 18}$ px NAC Craters | $N_{all}$ WAC Craters | NAC Image Difficult? | WAC Image Difficult? |
|---|---|---|---|---|---|---|
| Antonenko | 5.5[a], 7[b], 4[c] hrs | 5[a], 5.5[b] hrs | 636[a], 996[b], 653[c] | 949[a], 1308[b] | No | Yes |
| Chapman | 12[e] hrs | 4.5[5] hrs | 1197 | 644 | No | Highlands |
| Fassett | 5.7 hrs | 3 hrs | 814 | 1245 | Yes | No |
| Herrick | 8 hrs | 3 hrs | 808 | 336 | Yes | Yes |
| Kirchoff | 6 hrs | 7 hrs | 1027 | 1633 | No | No |
| Robbins | 2.4[d], 4.5[c] hrs | 1.4 hrs | 1060[d], 991[c] | 1114 | Yes | No |
| Singer | 3 hrs | 2 hrs | 652 | 1036 | No[f] | No |
| Zanetti | 7 hrs | 2.5 hrs | 930 | 1204 | Yes | No |

Data analysis conducted in/with: [a]Crater Helper Tools, [b]JMARS, [c]Moon Mappers, [d]Rim-tracing in ArcMap.

[e]Chapman also measured degradation state so this increased the time spent.

[f]Except for numerous heavily degraded craters.

**_Note to Typesetter:  Please leave the footnotes as letters, not numbers, because of the numerical values within the table.  (This was requested by a Reviewer.)_**

Table A2: Additional image manipulation performed by the different researchers using their preferred interface of choice.

| Researcher | Technique | Changed Pixel Scale (Zoom In/Out) | Contrast / Brightness Adjustment | Rotation |
|---|---|---|---|---|
| Antonenko[a] | Crater Helper Tools (ArcMap) and JMARS | X | X (WAC only) | [no] |
| Chapman | DS9 | X | X (WAC only) | X (NAC) |
| Fassett | CraterTools (ArcMap) | X | X | [no] |
| Herrick | Crater Helper Tools (ArcMap, NAC) and CraterTools (ArcMap, WAC) | X | X | [no] |
| Kirchoff | JMARS | X | X | [no] |
| Robbins[a] | Rim-Tracing (ArcMap) | X | X | [no] |
| Singer | Chord-Drawing (ArcMap) | X | X | [no] |
| Zanetti | CraterTools (ArcMap) | X | X | [no] |

[a]Also performed NAC counts with the Moon Mappers interface which did not allow these different manipulations.

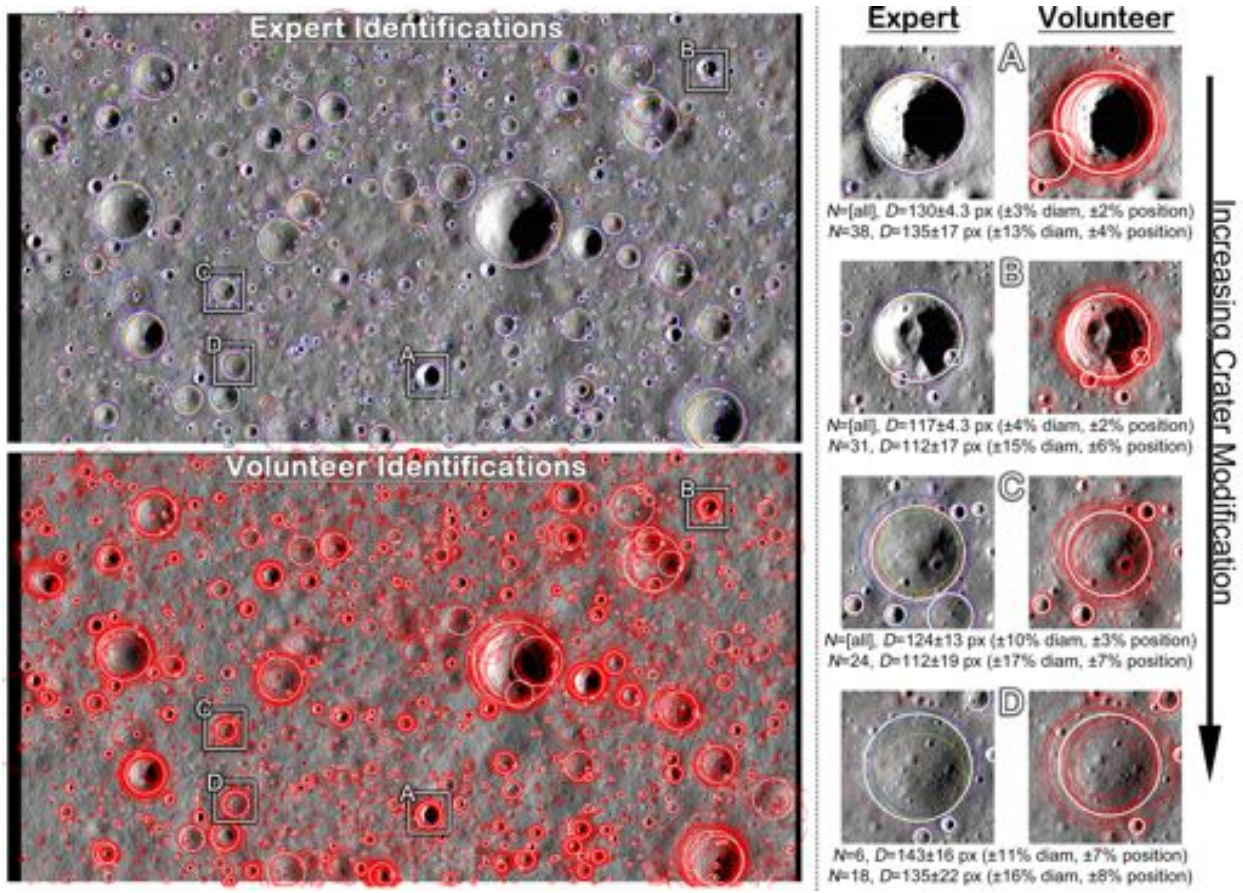

Figure 1: Left two panels are the full NAC areas analyzed in this study with markings overlaid. Top image shows expert markings, bottom shows volunteer data; both show craters only $D \geq 18$ px. The expert markings are color coded to correspond with the colors in Fig. 7. White, thicker circles are results from the clustering algorithm (see Section 2.3). On the right side, four example craters are shown in detail with expert markings and reduced craters (left column) and volunteer data and reduced craters (right column); the craters are in order of increasing modification/degradation with Class 1 at the top and Class 4 at the bottom (see Section 6 and Table 3). Captioned below each pair is the number ($N$) of persons who marked that crater and the mean diameter ($D$) with standard deviation. Values in parentheses are relative standard deviations (standard deviation of diameter divided by mean diameter; standard deviation of position divided by mean diameter). (Colors refer to electronic version. Print version shows individuals' craters as dark, thin circles and cluster results as light, thick circles.)

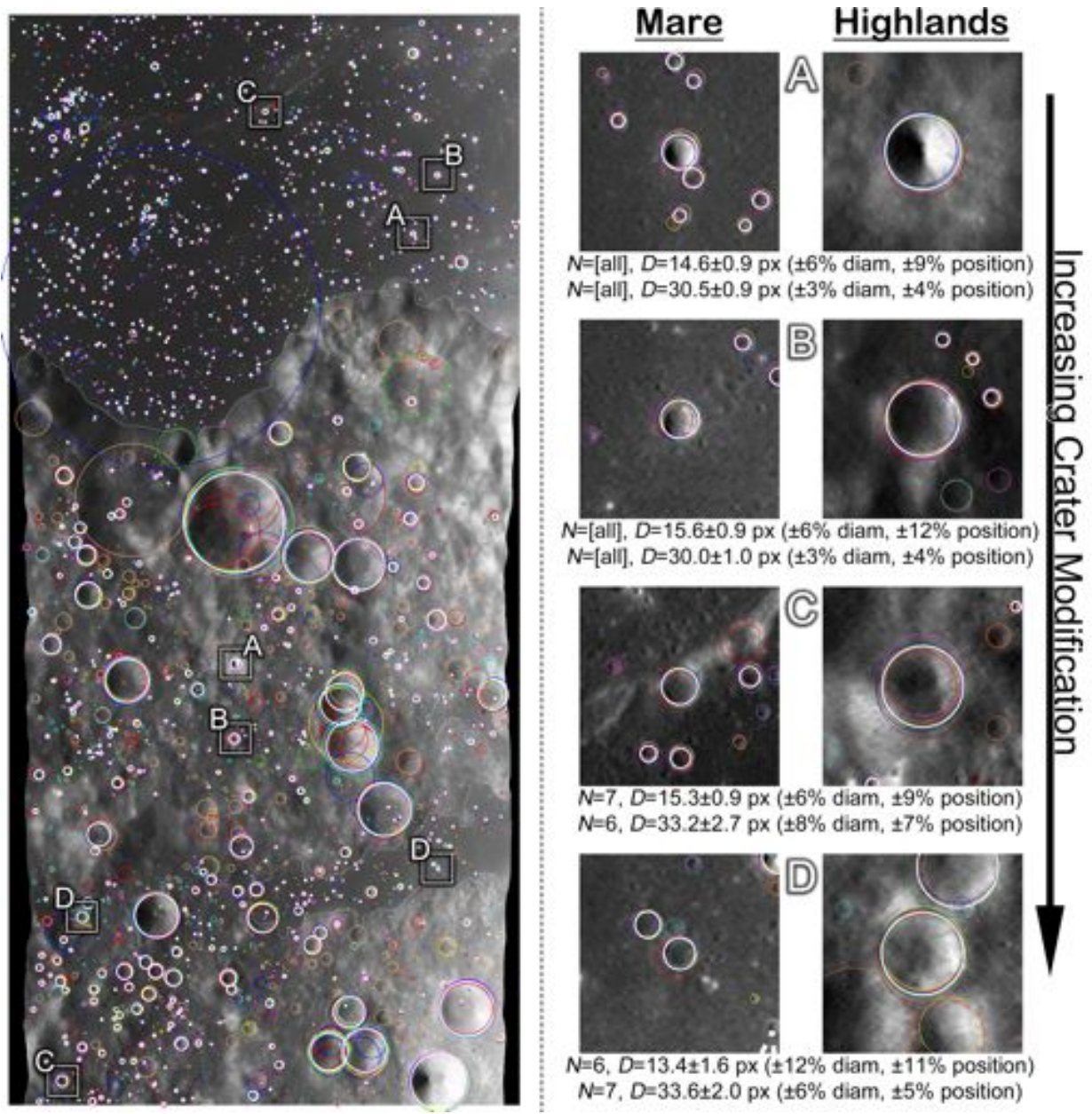

Figure 2: Left panel is the full WAC area analyzed in this study with markings overlaid (only $D \geq 7$ px craters are shown). The markings are color coded to correspond with the colors in Fig. 7, and white, thicker circles are results from the clustering algorithm. White dashed lines indicate boundaries between highlands and mare units. Right column follows Fig. 1 except are for mare (left) and highlands (right) craters as opposed to expert and volunteer, with data below each being mare (first line) and highlands (second line). (Colors refer to electronic version. Print version shows individuals' craters as dark, thin circles and cluster results as light, thick circles.)

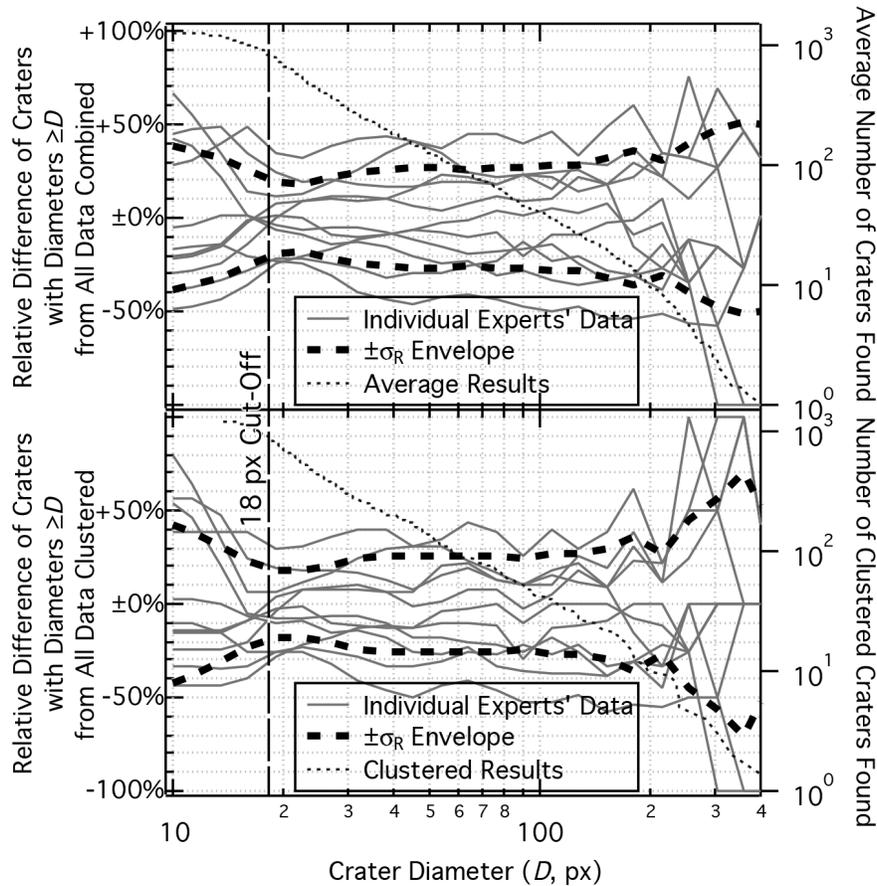

Figure 3: This figure shows experts' NAC data. Cumulative SFDs were made in multiplicative $2^{1/8}D$ bins for each expert's data. The relative differences of these results from the overall combined data (top) and clustered data (bottom) are shown here. The combined data is all experts' data added together and divided by the number of persons, while the clustered data represent the final results from the clustering code (see Section 3.1). The combined or clustered CSFDs were subtracted from each individual's CSFD, and these residuals were then divided by the combined or clustered CSFDs (thinner grey lines). The standard deviation was then calculated at each point to give a $\sigma_R$ envelope (thicker dashed lines). The vertical axis on the left indicates the percentage deviation, while the vertical axis on the right indicates the final number of craters found (thin dotted lines). Note that 18 px was the requested cut-off from each expert, so it is not unexpected that the results differ substantially below that point. This broadly illustrates that experts had a ~±30% dispersion in the number of craters measured that decreased to a ~±20% near the minimum crater diameter (maximum cumulative crater number).

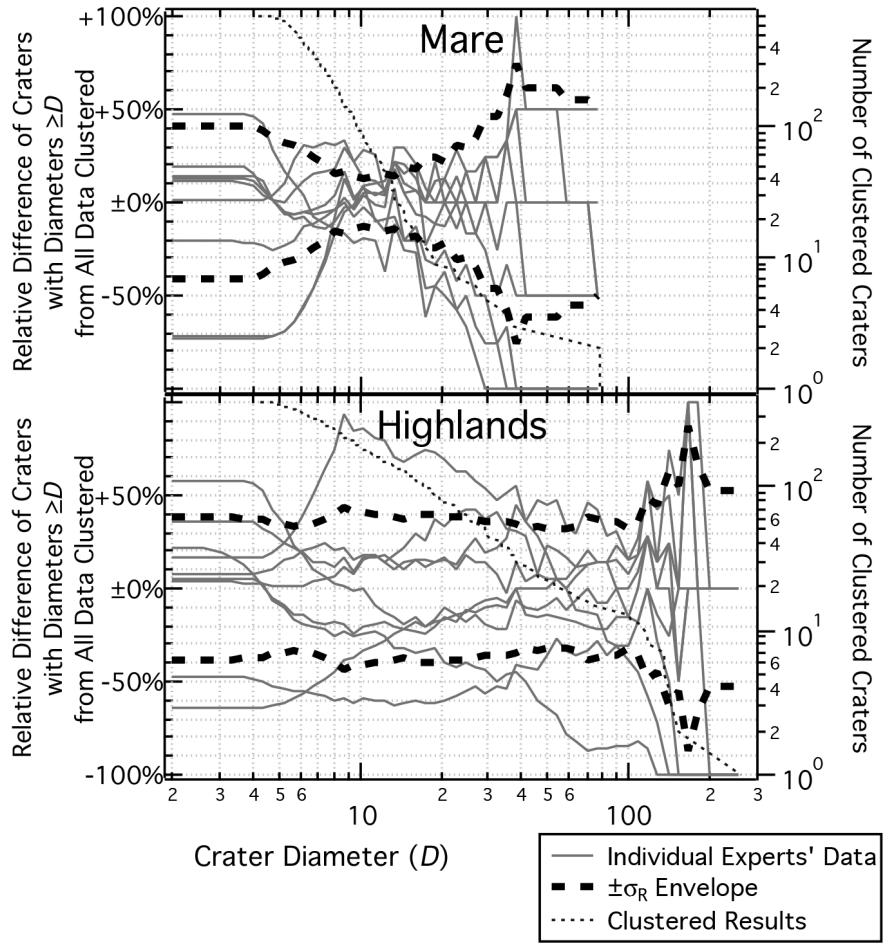

Figure 4: This figure shows experts' WAC data. It is the same as Fig. 3 (bottom) except for WAC data from mare region (top) and highlands (bottom). This illustrates that the mare and highlands regions behave very differently, where agreement for mare reaches a minimum dispersion of only ±10% for $D \approx 10$ px (below which various researchers were no longer complete in their counts) but the highlands have a dispersion in the number of craters found of ~35-45% almost independent of diameter and number of craters.

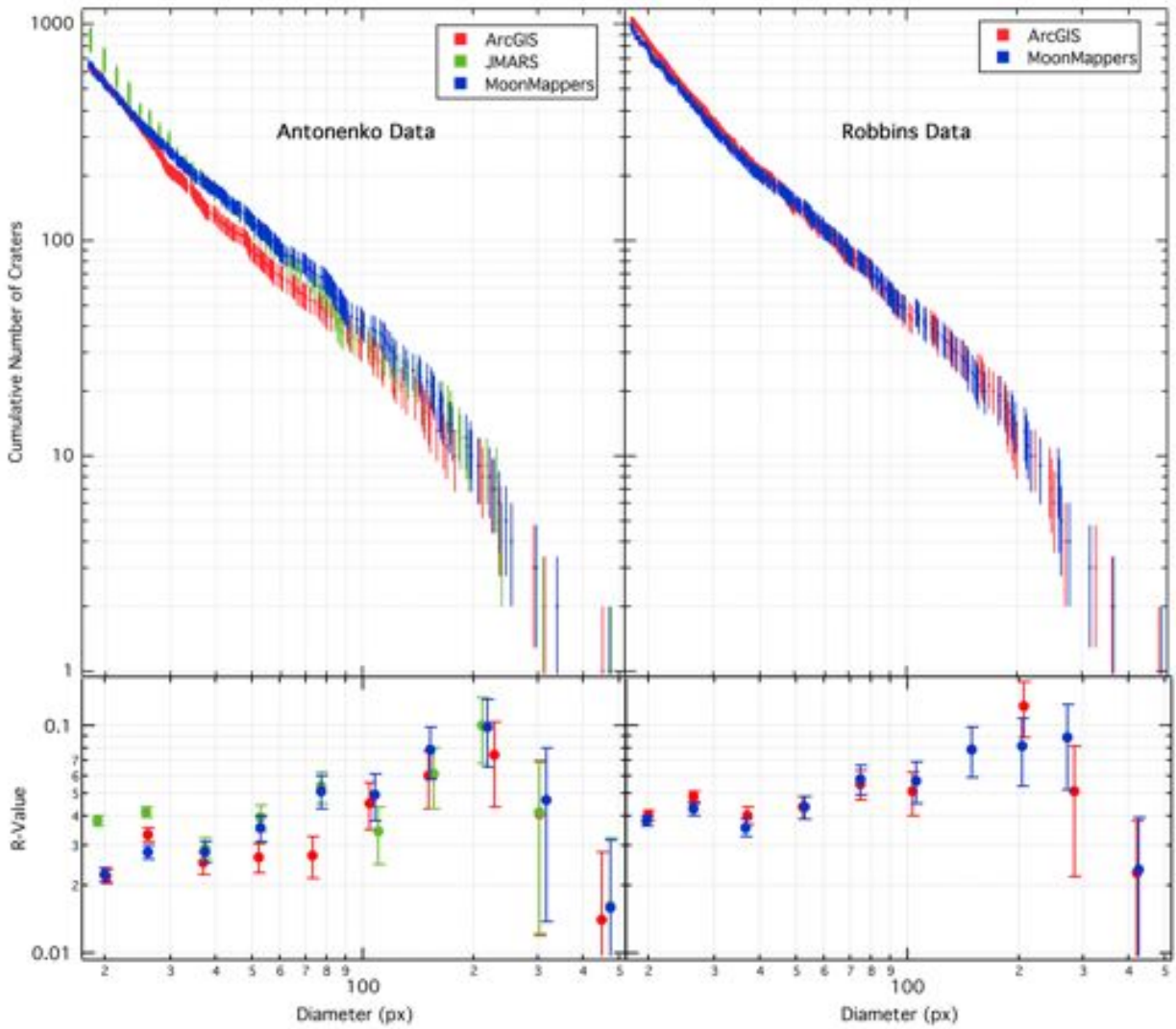

Figure 5: Cumulative SFDs (top) and R-plots (bottom) for both Antonenko (left) and Robbins (right) for the NAC image. CSFDs are unbinned histograms.

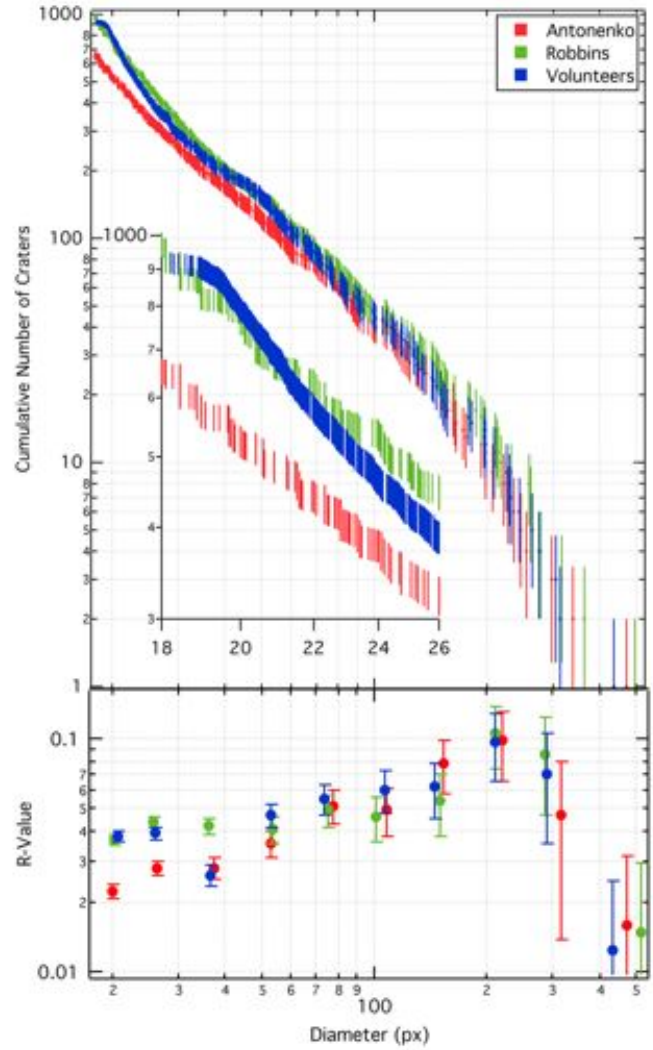

Figure 6: Same as Fig. 5 except with Antonenko's and Robbins' data from the Moon Mappers interface compared with volunteers' data. Inset focuses on the small diameters.

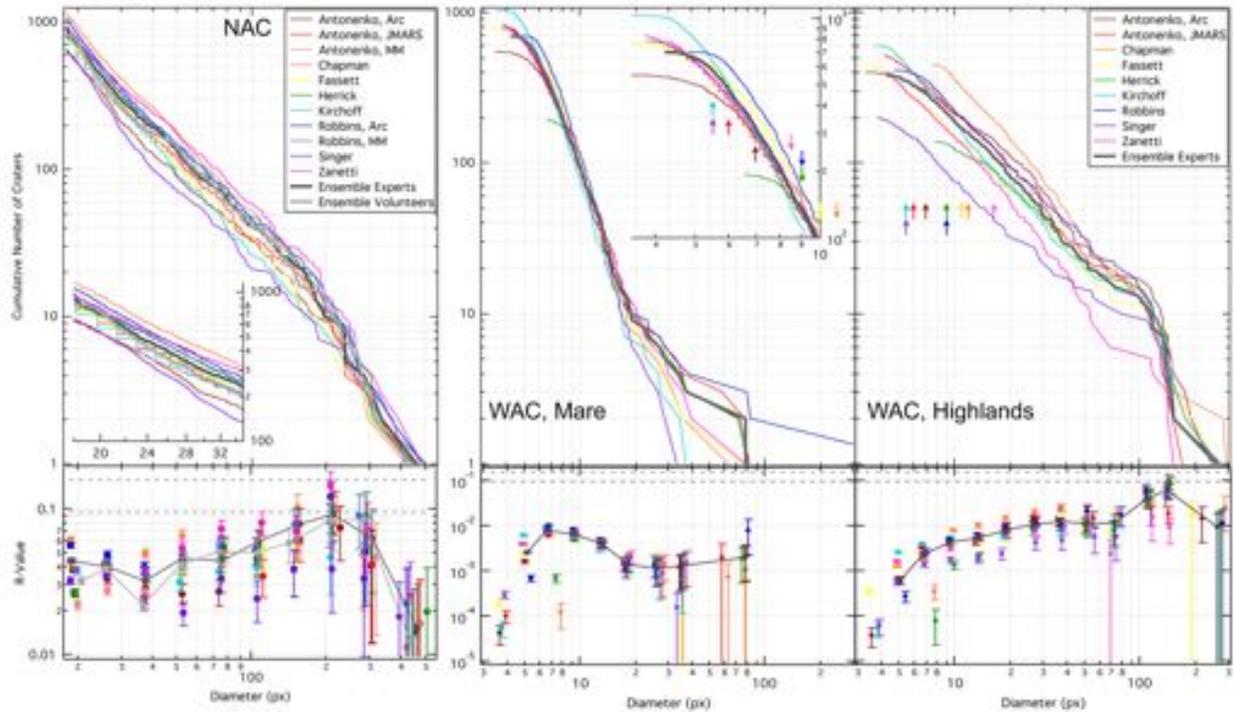

Figure 7: Cumulative size-frequency distributions (CSFDs) on the top row and R-plots on the bottom row of data for craters in the NAC image (left), mare in the WAC (middle), and highlands in the WAC (right). Colors correspond to different experts (see legend). Dark grey is the clustered expert data and light grey is the clustered volunteer data (latter is for NAC only). Dashed lines on R-plots correspond to 3% and 5% of geometric saturation. Inset in the NAC CSFD focuses on $18 \leq D \leq 35$ px; inset in the mare WAC CSFDs focuses on $3.5 \leq D \leq 10$ px, and small vertical arrows correspond to where each expert estimated their completeness to be. Horizontal and vertical axes are different for the NAC versus WAC columns because of different completeness levels. Error bars have been removed from the CSFDs for clarity; since the vertical scale is the cumulative number of craters, uncertainty would be $N^{1/2}$ (e.g., ±10 for $N_{\text{cumulative}} = 100$ and ±32 for $N_{\text{cumulative}} = 1000$).

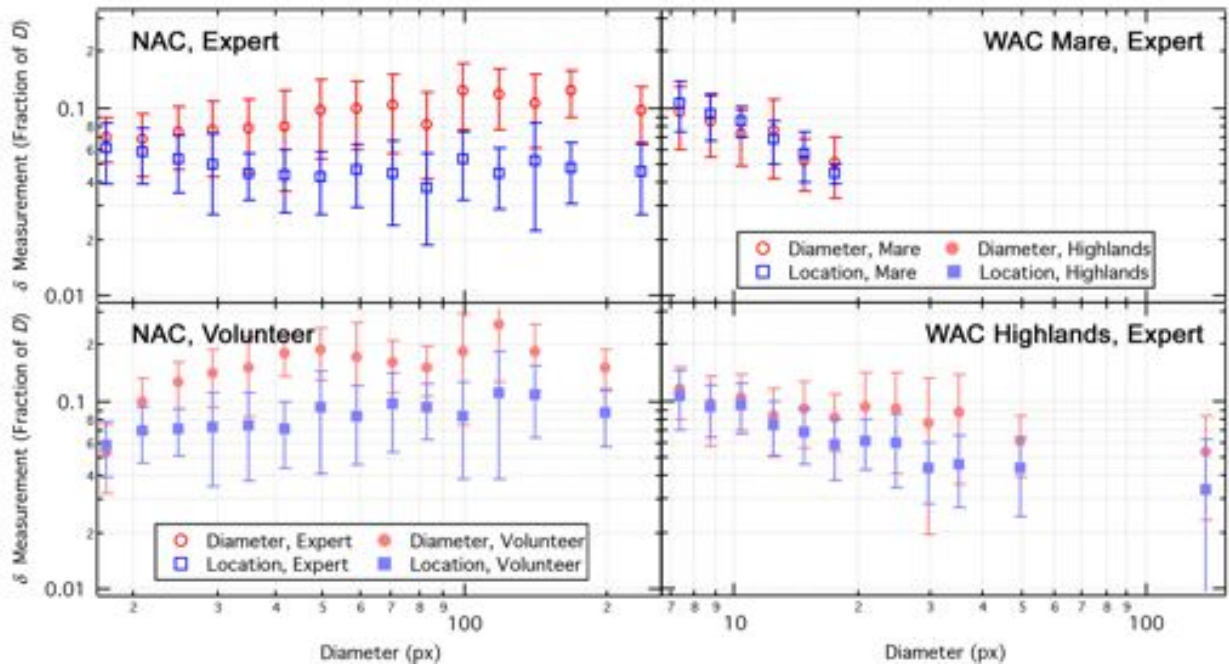

Figure 8: Standard deviation of the reduced results for crater diameter and location, as a function of diameter, where standard deviation is expressed as a fraction of crater diameter (location in this case is the average of $\delta x$ and $\delta y$ values). This means that $\delta D/D$ and $(\delta x + \delta y)/(2D)$ are plotted on the ordinate, so they are not in values of pixels; this acts to remove the linear function of $D$ trend from the data. Craters are binned in $2^{1/2}D$ multiplicative bins and the mean and standard deviation of the $\delta D$ and $0.5 \cdot (\delta x + \delta y)$ are shown. Left panel is NAC, right panel is WAC (WAC data are experts only). Note that the horizontal axes are different because the NAC data were truncated at a minimum 18 pixels while WAC data here were truncated at 7 pixels.

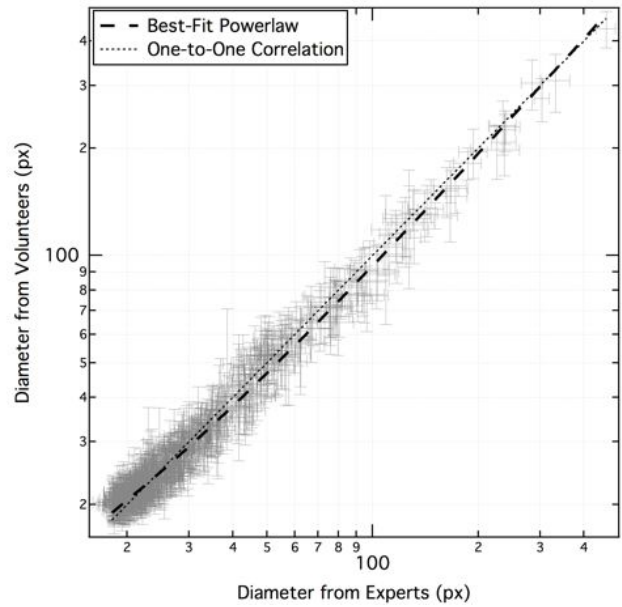

Figure 9: 699 craters $D \geq 18$ px were matched between the reduced expert and volunteer catalogs and their diameters are displayed versus each other on this graph with corresponding $1\sigma$ uncertainties from the means of the individual markings that went into each reduced crater. Ideally, all markings should be on or randomly scattered about the dotted line. A small deviation is observed with the parameters $D_{volunteer} = 5.3 + 0.58 \cdot (D_{expert})^{1.1}$, but it is statistically identical to a 1:1 line ($D_{volunteer} = D_{expert}$).

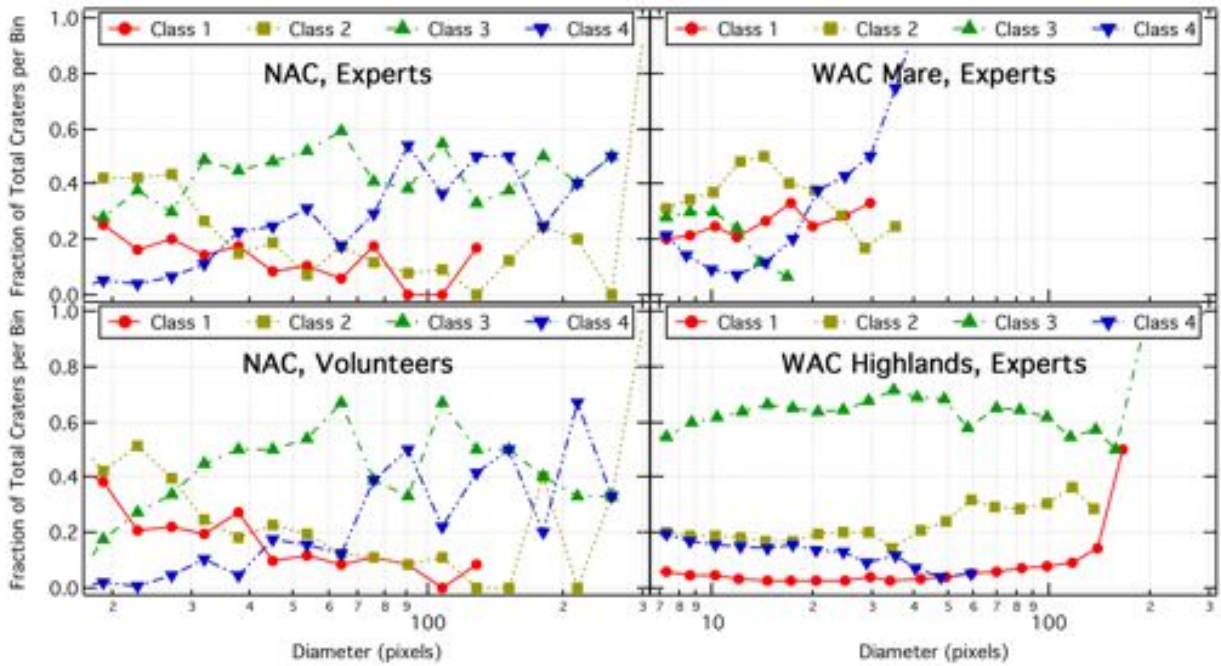

Figure 10: Left— Fraction of craters per $2^{1/4}D$ diameter bin found by experts (top) and volunteers (bottom) in the NAC image, separated by preservation state. Right— Same as left except for the WAC image's mare region (top) and highlands region (bottom). Note that the horizontal axes are different because of different completeness limits.

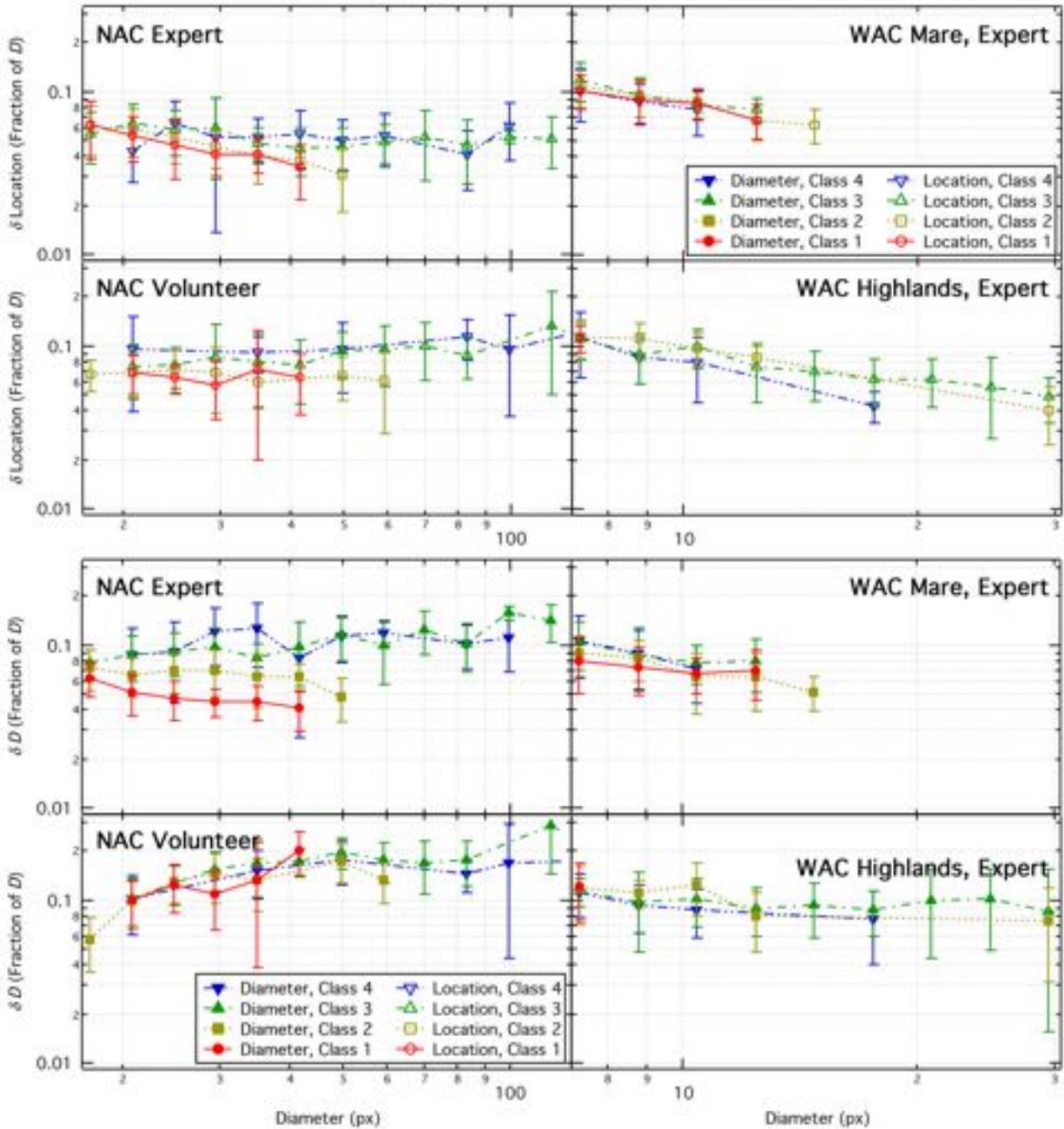

Figure 11: Similar to Fig. 8, but this Figure expands those data by preservation class. The left column are NAC data, right column are WAC data. Each set of four are the same category of data (persons, area/data) but the top group are locations and bottom group are diameters. In each set of four, the top-left are experts and bottom-left are volunteers. WAC data have been divided by mare (top right) and highlands (bottom right); the diameter range has been limited for the WAC data because all larger bins have fewer than 5 craters in them. Note that the horizontal

axes are different between the two columns because of different completeness limits.